\documentclass[reprint,superscriptaddress,aps,prx,floatfix, nofootinbib]{revtex4-2}
\usepackage{microtype}
\usepackage[utf8]{inputenc}
\usepackage[tbtags]{amsmath}
\allowdisplaybreaks
\usepackage{amsthm,amssymb,dsfont,bbm,xfrac,tabularx,booktabs}
\usepackage[linesnumbered,ruled,vlined]{algorithm2e}
\usepackage{upgreek}
\usepackage{mathtools}
\DeclareMathAlphabet{\mathcal}{OMS}{cmsy}{m}{n}
\usepackage{comment}
\usepackage{subcaption}
\usepackage{caption}
\newcommand{\poly}{\mathrm{poly}}

\makeatletter
\usepackage{color}
\usepackage{babel}
\usepackage{units}

\usepackage{tikz}
\usetikzlibrary{positioning}

\tikzset{%
	every neuron/.style={
		circle,
		draw,
		minimum size=1cm
	},
	neuron missing/.style={
		draw=none, 
		scale=4,
		text height=0.333cm,
		execute at begin node=\color{black}$\vdots$
	},
}

\usepackage{color}
\usepackage{xcolor}
\usepackage{dsfont}

\usepackage[cal=boondoxo]{mathalfa}
\usepackage{grffile}

\usepackage[unicode=true,pdfusetitle,
bookmarks=true,bookmarksnumbered=false,bookmarksopen=false,
breaklinks=false,pdfborder={0 0 1},backref=false,colorlinks=true]
{hyperref}
\hypersetup{pdfborderstyle=,urlcolor=orange,linkcolor=blue,citecolor=red}

\usepackage[capitalise]{cleveref}

\catcode`,\active

\catcode`\,12

\newsavebox{\@brx}
\newcommand{\llangle}[1][]{\savebox{\@brx}{\(\m@th{#1\langle}\)}%
	\mathopen{\copy\@brx\kern-0.5\wd\@brx\usebox{\@brx}}}
\newcommand{\rrangle}[1][]{\savebox{\@brx}{\(\m@th{#1\rangle}\)}%
	\mathclose{\copy\@brx\kern-0.5\wd\@brx\usebox{\@brx}}}

\makeatletter
\def\maketitle{
	\@author@finish
	\title@column\titleblock@produce
	\suppressfloats[t]}
\makeatother

\def\ddefloop#1{\ifx\ddefloop#1\else\ddef{#1}\expandafter\ddefloop\fi}
\def\ddef#1{\expandafter\def\csname bb#1\endcsname{\ensuremath{\mathbb{#1}}}}
\ddefloop ABCDEFGHIJKLMNOPQRSTUVWXYZ\ddefloop
\def\ddef#1{\expandafter\def\csname sf#1\endcsname{\ensuremath{\mathsf{#1}}}}
\ddefloop ABCDEFGHIJKLMNOPQRSTUVWXYZ\ddefloop
\def\ddef#1{\expandafter\def\csname c#1\endcsname{\ensuremath{\mathcal{#1}}}}
\ddefloop ABCDEFGHIJKLMNOPQRSTUVWXYZ\ddefloop
\def\ddef#1{\expandafter\def\csname vec#1\endcsname{\ensuremath{\mathbf{#1}}}}
\ddefloop abcdefghijklmnopqrstuvwxyz\ddefloop
\def\ddef#1{\expandafter\def\csname mat#1\endcsname{\ensuremath{\mathbf{#1}}}}
\ddefloop ABCDEFGHIJKLMNOPQRSTUVWXYZ\ddefloop

\allowdisplaybreaks

\makeatother

\begin{document}
	\title[]{Are Neural Networks Collision Resistant?}

	\author{Marco Benedetti}
	\affiliation{Department of Computing Sciences, Bocconi University, Milano, Italy}
	
	\author{Andrej Bogdanov}
	\affiliation{University of Ottawa, Canada}
	
	\author{Enrico M. Malatesta}
	\affiliation{Department of Computing Sciences, Bocconi University, Milano, Italy}
	
	\author{Marc M\'{e}zard}
	\affiliation{Department of Computing Sciences, Bocconi University, Milano, Italy}
	
	\author{Gianmarco Perrupato}
	\affiliation{Department of Computing Sciences, Bocconi University, Milano, Italy}
	
	\author{Alon Rosen}
	\affiliation{Department of Computing Sciences, Bocconi University, Milano, Italy}
	
	\author{Nikolaj I. Schwartzbach}
	\affiliation{Department of Computing Sciences, Bocconi University, Milano, Italy}
	
	\author{Riccardo Zecchina}
	\affiliation{Department of Computing Sciences, Bocconi University, Milano, Italy}

	\begin{abstract}
		When neural networks are trained to classify a dataset, one finds a set of weights from which the network produces a label for each data point. We study the algorithmic complexity of finding a collision in a single-layer neural net, where a collision is defined as two distinct sets of weights that assign the same labels to all data. For binary perceptrons with oscillating activation functions, we establish the emergence of an overlap gap property in the space of collisions. This is a topological property believed to be a barrier to the performance of efficient algorithms. The hardness is supported by numerical experiments using approximate message passing algorithms, for which the algorithms stop working well below the value predicted by our analysis. Neural networks provide a new category of candidate collision resistant functions, which for some parameter setting depart from constructions based on lattices. Beyond relevance to cryptography, our work uncovers new forms of computational hardness emerging in large neural networks which may be of independent interest.
	\end{abstract}
	\maketitle

	Neural networks are models of computation at the core of modern machine learning. In feed-forward networks used as classifiers, given an input vector of size $N$, the network computes an output by alternating layers of linear transformations based on `weight' matrices with non-linear activation functions. Arguably, the simplest neural network is the \emph{binary perceptron} \cite{mcculloch1943logical,cover1965geometrical}, consisting of a single layer and binary weights $\vecx \in \{-1,1\}^N$. Given a matrix of $P$ data points in $N$-dimensional space, $\matA \in \bbR^{P \times N}$ (also called the \emph{disorder}),
	the perceptron outputs the labels $f_\matA(\vecx) = {\varphi}(\matA \vecx)$, where the activation function	$\varphi : \bbR \rightarrow \{-1,1\}$ is applied element-wise. 

	Here we consider the perceptron from a dual point of view. Given the disorder $\matA$, we use as inputs the weights $\vecx$ and study the function $f_\matA(\vecx) $ in a regime of large dimensions where $N,P\to\infty $ with fixed $\alpha=P/N$. Several constraint satisfaction problems (CSPs) arise in this context, listed in order of non-increasing algorithmic difficulty: 
	\begin{itemize}
		\item {\em inversion:} given a matrix $\matA \in \bbR^{P \times N}$ and a vector of labels $\boldsymbol{\ell} \in \{-1,1\}^P$,  find any set of inputs $\vecx\in\{-1,1\}^N$ such that $f_\matA(\vecx)=\boldsymbol{\ell}$.
		\item {\em second-preimage:}  given a matrix $\matA \in \bbR^{P \times N}$, and a 
		`teacher' $\vecx \in \{-1,1\}^N$ generating labels  $\boldsymbol{\ell}=f_\matA(\vecx) \in \{-1,1\}^P$, find any `student' $\vecx'\in \{-1,1\}^N$ with $\vecx' \ne \vecx$ such that $f_\matA(\vecx')=\boldsymbol{\ell}$.
		\item {\em collision-finding:} given a matrix $\matA \in \bbR^{P \times N}$, find any two inputs $\vecx\neq \vecx'\in \{-1,1\}^N$ such that $f_\matA(\vecx)=f_\matA(\vecx')$.
	\end{itemize}
	We are interested in the typical case complexity of CSPs, where the matrix $\matA$  has random identically distributed independent (i.i.d.) entries, and also the labels are i.i.d. random in the inversion problem. 
	Statistical physics has a long tradition of studying random CSPs, such as inversion and second-preimage variants \cite{kirkpatrickselman1994,mezard2002analytic,krzakala2007gibbs,mezard2009information}, yet the collision-finding problem itself had not been addressed within this framework until now. Notice the difference between second-preimage and collision-finding: in the latter, there is full freedom to choose both $\vecx$ and $\vecx'$ in a way that might depend on $\matA$, while in the former one of the two is fixed. A function $f$ for which there are no polynomial time algorithms that find collisions\footnote{Formally, collision resistance is defined as: for every efficient algorithm $C(\cdot)$ and any constant $c>0$, the probability that $C(\matA)$ outputs $\vecx \neq \vecy$ with $f_\matA(\vecx) = f_\matA(\vecy)$ is smaller than $N^{-c}$ for any sufficiently large $N$ (where the randomness is taken over the coins used by $C$)} is said to be collision resistant. 
	
	The collision-finding problem, on the other hand, has been extensively studied in cryptography. A function for which the collision-finding problem is hard is said to be \emph{collision resistant}. If in addition, the function is \emph{shrinking}, i.e.\ the output size $P$ is smaller than the input size $N$, it is known as a \emph{collision resistant hash function} (CRH). CRHs are of fundamental importance and form the basis of many cryptographic protocols and security guarantees. In fact, they are an integral part to securing data privacy \cite{goldwasser2019knowledge}, in applications including commitment schemes \cite{blum1983coin}, encryption \cite{diffie2022new} and secure computation \cite{yao1982protocols}. In this paper, our main goal is to understand whether the collision-finding problem in neural networks with a suitably chosen activation function is algorithmically hard. 
	
	Statistical physics studies on random Boolean satisfiability~\cite{mezard2002analytic,mezard2002random,mezard2005clustering,monasson1999determining} have linked typical-case algorithmic hardness to the presence of peculiar geometrical properties in the space of solutions of CSPs. Building upon this geometric approach~\cite{Zdeborova2007coloring,baldassi2015subdominant,Baldassi2022learning,Baldassi2023}, it has been rigorously shown that the presence of a disconnectivity feature known as the overlap gap property (OGP) in the solution space of CSPs leads to the failure of stable algorithms~\cite{gamarnik2021overlap,gamarnik2025turing} and larger classes including online algorithms and low-degree polynomials. In many CSPs, the most efficient solvers identified to date belong to this class, giving rise to the conjecture that OGP is a marker of computational hardness\footnote{CSPs with a linear structure are considered somewhat an exception to this hardness conjecture, since despite the disconnectivity properties of their space of solutions, they can be always solved in polynomial time by Gaussian elimination.}.

	Examples are the symmetric binary perceptron problem~\cite{gamarnik2022algorithms}, number partitioning \cite{gamarnik2023algorithmic}, maximum independent set on random graphs \cite{gamarnik2025optimal}, and the optimization of $p$-spin glass Hamiltonians \cite{gamarnik2021overlapPspin,gamarnik2024hardness}. 

	Our main result is the identification of a novel candidate for collision resistant hash function based on a family of neural networks, where the OGP is used as a criterion for collision resistance between pairs of inputs of large Hamming distance. We then argue that by applying an error-correcting code upstream of a perceptron with an oscillating activation one obtains a still shrinking and collision resistant function. 
	
	These new candidate CRHs are markedly different from existing constructions from e.g.\ lattices, and hint at a new source of hardness for use in cryptography.
	
	\section{Summary of Our Contributions}
	
	We study here the collision resistance of a family of perceptrons, where the activation function is a periodic square wave\footnote{As shown in the appendices, our analysis holds for a general $\varphi$. We concentrate mostly on the family of periodic square-wave activations merely because of their potential relevance in cryptography.}. Specifically, we consider the following activation function, called the Square Wave Perceptron (SWP):
	\begin{equation}
		\varphi_ {\delta}(h) = \text{sgn}\left(\frac{h}{2\delta} \,\mathrm{mod}\,1\right)\,,
		\label{eq:defSWP}
	\end{equation}
	where $x \,\mathrm{mod}\,1$ is the unique real number in $[-\frac12, \frac12)$ that differs from $x$ by an integer, and $\delta \in \mathbb{R}^+$ is a parameter that controls the oscillation period. The function (\ref{eq:defSWP}) is a square-wave with period $2\delta$. The choice of activation is ultimately guided by the goal of obtaining a shrinking collision resistant function which, as we show, is crucially dependent on $\delta$. 
	
	Some properties of the SWP have been studied in \cite{benedetti2025overlap} in average-case variants of inversion and second-preimage problems. Here we focus on the collision problem. In  Section \ref{sec::existence_collisions}, we study the existence of collisions in the large size limit, in the regime where the inputs are at an extensive Hamming distance, or equivalently, when the overlap  
	$q_1=\vecx^{\top}\vecy/N$ between the two inputs $\vecx, \vecy\in \{-1, 1\}^N$ forming a collision is strictly smaller than one, $|q_1|\leq 1-\Omega(1)$. We show that above a certain threshold of $\alpha$, depending on $q_1$ and $\delta$, with high probability collisions do not exist. The value of this threshold computed by using the replica method with a replica-symmetric Ansatz \cite{mezard1987spin} coincides with a first-moment computation, leading us to conjecture that the replica-symmetric estimate coincides with the actual SAT/UNSAT transition in the collision problem. This conjecture is supported by numerical simulations based on exhaustive search of collisions on finite-size systems. 
	
	In Section \ref{sec::ogp_upper_bound} we move to the main focus of the paper, namely determining if there is a region of $\alpha$ where collisions exist but are hard to find. Using the first moment method we show that, conditioning to $|q_1|<1$ the space of solutions of the collision-finding problem presents an OGP, which implies the failure of stable algorithms. We conjecture that in this regime, finding collisions with $|q_1|<1$ is infeasible. 
	
	By tuning the parameter $\delta$, we show that the conjectured hard region extends to values of $\alpha$ smaller than one, which is a request to construct a hash function. More precisely, we show that there exist $\alpha_{*},\delta_*$ and $q_*<1$ such that for $\delta<\delta_*$ the space of collisions with overlap $q_1< q_*$ exhibits OGP for $\alpha_{\star}<\alpha<1$. Still, a solver could target collisions with overlap $q_*<q<1$, where our analysis does not imply collision resistance for $\alpha<1$. To remedy this, in Section \ref{sec::HFNN} we propose a collision resistent hash function obtained by composing an error correcting code with the square wave perceptron. The error correcting code removes collisions with overlaps in the region $q_*<q<1$, leading to a bona-fide collision resistant hash function.
	
	In Section \ref{sec:AlgAtt} we provide some numerical evidence supporting our conjectured CRHs. We study numerically two algorithmic attacks to find collisions, showing that both of them stop working at values of $\alpha$ smaller than the value $\alpha_*$ where we predict the presence of an OGP.

	\section{Existence of Collisions} 
	\label{sec::existence_collisions}
	
	Consider the number of collisions with internal overlap $q_1$:
	\begin{equation}
		Z (q_1; \boldsymbol{A}) := \sum_{\boldsymbol{c}\in \{\pm 1\}^{2N}} \mathbb{X}_{\boldsymbol{A}}(\boldsymbol{c}) \, \delta\left(q_i(\boldsymbol{c}) - q_1 \right), 
		\label{eq:number_of_collisions}
	\end{equation}
	where $\mathbb{X}_{\boldsymbol{A}}(\boldsymbol{c})$ is an indicator function, equal to one if the pair of inputs $\boldsymbol{c} = (\vecx,\vecy)$ forms a collision and zero otherwise:
	\begin{equation}
		\mathbb{X}_{\matA}(\boldsymbol{c}) = \prod_{\mu=1}^P\Theta\left[f_\matA(\vecx)_{\mu} f_\matA(\vecy)_{\mu} \right].
	\end{equation}
	In this notation $\Theta(\cdot)$ is the Heaviside step function, and $f_\matA(\cdot)_{\mu}$ the $\mu$-th element of vector $f_\matA(\cdot)$. 
	We have also defined the `\emph{internal}' overlap of a collision $\boldsymbol{c}$ as the normalized scalar product between the colliding inputs 
	\begin{equation}
		\label{eq::internal_overlap}
		q_i(\boldsymbol{c}) =  \frac{1}{N} \vecx^\top \vecy \,.
	\end{equation}

	In the limit $N, P \to \infty$ with $\alpha = P/N$ fixed, the random variable $1/N\log Z$ concentrates on its average value, the so called $\emph{quenched}$ free entropy~\cite{talagrand1999}:

	\begin{equation}
		\Phi(q_1)= \lim_{N\to \infty} \frac{1}{N} \mathbb{E}_{\boldsymbol{A}}\log Z (q_1 ; \boldsymbol{A}).
		\label{eq::quenched_free_entropy_existence}    
	\end{equation}
	Hence, $Z (q_1 ; \boldsymbol{A})= \exp(N\Phi(q_1)+o(N))$, and if $\Phi<0$ one has $Z (\boldsymbol{A})=0$ with high probability in the large $N$ limit. Therefore, collisions with internal overlap $q_1$ exist up to a threshold $\alpha_c(q_1)$ where $\Phi=0$. 
	We estimated $\alpha_c(q_1)$ using two complementary approaches. First, we derived an upper bound $\alpha^a_c(q_1)$ to $\alpha_c(q_1)$ by using the first moment method. Because of Markov's inequality, the probability that the random variable $Z (q_1; \boldsymbol{A}) \ge  1$ is bounded by
	\begin{equation}
		P\left[Z (q_1; \boldsymbol{A}) \ge 1 \right] \le \mathbb{E}_{\boldsymbol{A}} Z (q_1; \boldsymbol{A}) = e^{N \Phi^\mathrm{a}(q_1)},
	\end{equation}
	where the \emph{annealed} free entropy $\Phi^\mathrm{a}(q_1)$ is given by
	\begin{equation}
		\Phi^{\mathrm{a}}(q_1)= \lim_{N\to \infty} \frac{1}{N} \log \mathbb{E}_{\boldsymbol{A}} Z (q_1 ; \boldsymbol{A})\,.
		\label{eq::annealed_free_entropy_existence}    
	\end{equation}

	The value of $\alpha$ for which the annealed free entropy vanishes therefore gives an upper bound to $\alpha_c(q_1)$. As we show in the Supplementary material, this is given by
	\begin{equation}
		\alpha_c^\mathrm{a}(q_1) = - \frac{\log(2) + H_B(q_1)}{\log  \int Dz  \, F_\varphi \left(\sqrt{q_1} z; \sqrt{1-q_1}\right)},
	\end{equation}
	where
	\begin{equation}
		H_B(q_1) \equiv -\frac{1-q_1}{2} \log \left(\frac{1-q_1}{2} \right) - \frac{1+q_1}{2} \log \left(\frac{1+q_1}{2}\right),
	\end{equation}
	\begin{equation}
		F_\varphi(x; \sigma) \equiv 1 - 2I_\varphi(x; \sigma) \left( 1 - I_\varphi(x; \sigma)\right),
	\end{equation}
	\begin{equation}
		I_\varphi(x; \sigma) \equiv \int Dh \, \Theta\left[ \varphi\left(\sigma h + x \right)\right].
	\end{equation}

	Second, we applied the replica method \cite{mezard1987spin} to compute the Replica Symmetric (RS) upper bound $\Phi^{RS} \ge \Phi$ to the quenched free entropy~\eqref{eq::quenched_free_entropy_existence}, and the corresponding upper bound $\alpha_c^{RS}\ge \alpha_c$. Details can be found in the Supplementary Material. 
	Here we mention that previous applications of the replica method using the RS approximation
	~\cite{krauth1989storage} have already been shown to give the correct result for the storage capacity in inversion problems~\cite{ding2019capacity,huang2024capacity} with binary inputs, whereas for continuous inputs one needs Replica Symmetry Breaking~\cite{barkai1992broken,engel1992storage,baldassi2019_relu,Annesi2025}.
	
	Interestingly, we found that the more refined RS computation of the quenched free entropy~\eqref{eq::quenched_free_entropy_existence}, does not change the result found using the first moment method $\Phi^{\mathrm{RS}}(q_1) = \Phi^{\mathrm{a}}(q_1)$. This leads also to the same prediction for the value of $\alpha$ up to which collision exist $\alpha_c^{\mathrm{RS}}(q_1) = \alpha_c^{\mathrm{a}}(q_1)$, suggesting that the annealed bound is tight and, in this case, exact. \\
	
\begin{figure}[ht]
		\centering
		\includegraphics[trim={0 0 0 1.9cm}, clip, width=1.0\linewidth]{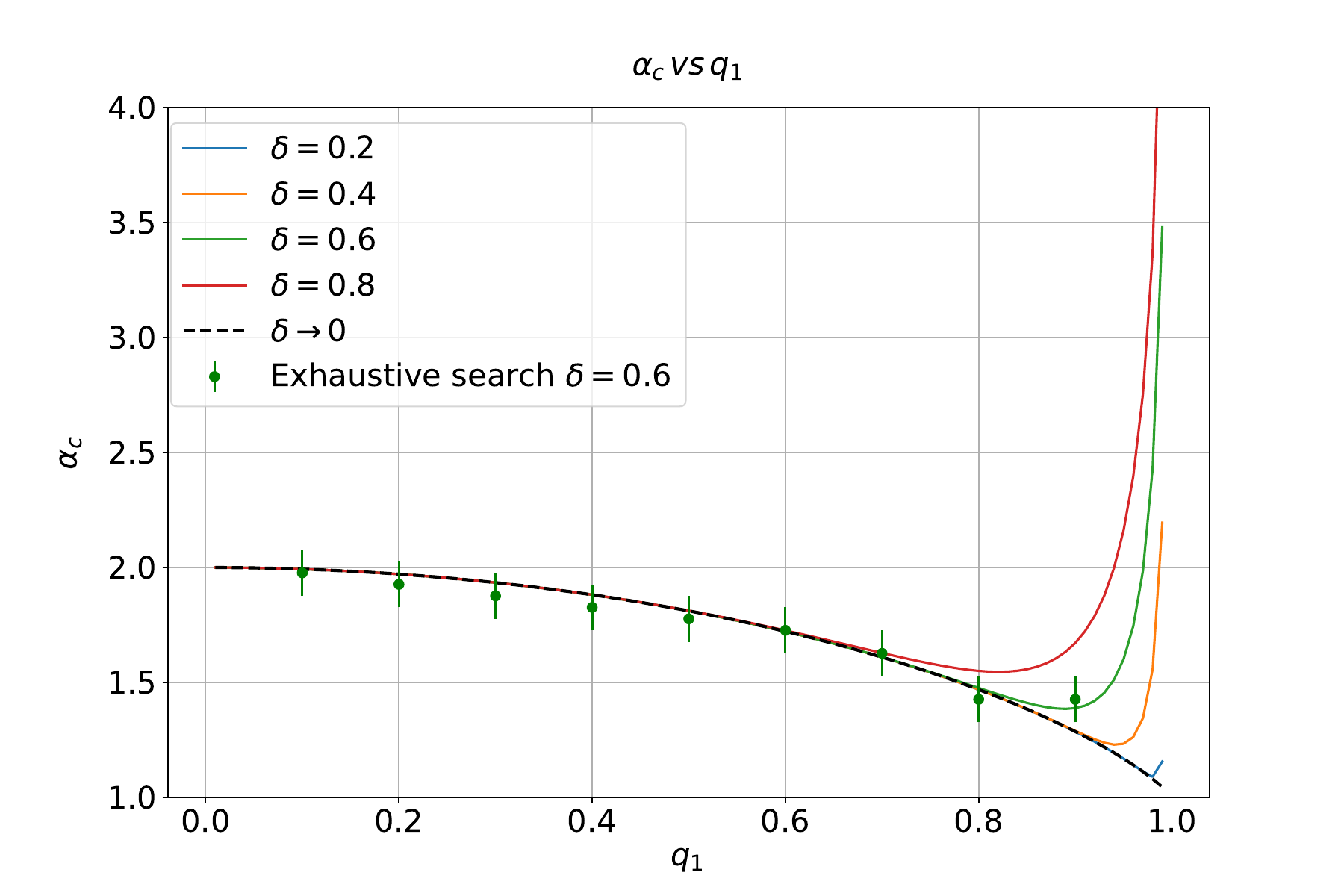}
		\caption{Full lines: analytic prediction for the transition $\alpha_c$ above which no collision exists versus the internal overlap $q_1$, for different values of $\delta$. Dots: numerical estimates from exhaustive search.}
		\label{fig:sat_unsat}
	\end{figure}

	\Cref{fig:sat_unsat} displays $\alpha_c^{\mathrm{RS}}$ as a function of $q_1$ for various values of $\delta$. For $\alpha>\alpha_c^{RS}(q_1)$ the number of collisions is exponentially small. For all $\delta$ strictly larger than zero, $\alpha_c(q_1)$ diverges in the limit $q_1\to 1$ as $\sim (1-q_1)^{-1/2}$. This is to be expected: trivial collisions with $q_1=1$, i.e., $\vecx=\vecy$ are always present. The divergence around $q_1\sim 1$ means that sub-extensive collisions exist for all $\alpha=O(1)$ and independent of the activation function. In the SM we show that the computation of the annealed free entropy for $q_1\approx 1-1/N$  consistently predicts $\alpha^{\mathrm{a}}_c\propto N^{1/2}$. 
	
	Finally, notice that for each value of $q_1$, decreasing $\delta$ leads to a smaller value of $\alpha_c^{RS}$. In particular in the small $\delta$ limit, which we call \emph{strong hashing limit}, one obtains 
	\begin{equation}
		\alpha_c^\mathrm{a}(q_1) = 1 + \frac{H_B(q_1)}{\log 2}, 
	\end{equation}
	which is a monotonically decreasing function of $q_1$. This curve is shown in black dashed in~\Cref{fig:sat_unsat}. Dots in Fig.~\ref{fig:sat_unsat} present a numerical estimate of $\alpha_c$, obtained by exhaustively searching the space of weights and enumerating how many collisions exist with a given value of $q_1$ (see 
	the SM for details). The procedure takes an exponential time in $N$, so the estimate is based on data with small $N=14, 16, ... ,24$. Still, finite size effects are small, and the extrapolation are in excellent agreement with the analytic prediction.

	\section{Algorithmic Barriers to Collision-Finding} 
	\label{sec::ogp_upper_bound}
	
	\subsection{Multi Overlap Gap Property.} 
	\label{sec:m_OGP}
	The overlap gap property is a geometric property of the solution space of CSPs, which is proven to imply the failure of stable algorithms. Informally, the stability of an algorithm $\mathcal{A}$, defined as a map from the instance space to the solution space of a CSP, requires that a small perturbation of the input results in a small perturbation of the output. In various CSPs, the best known solvers have been shown to belong to this class, fostering the belief that the presence of OGP is a signature of general algorithmic hardness \cite{gamarnik2021overlap}. Examples of algorithms proven to satisfy this stability property are the Kim-Roche algorithm for the symmetric binary perceptron \cite{gamarnik2022algorithms}, approximate message passing (AMP) \cite{gamarnik2021overlapPspin}, low-degree polynomials, Langevin dynamics and low-depth Boolean circuits \cite{gamarnik2024hardness,Wein_2022}.\\
	
	In this section we discuss a variant of OGP called multi overlap gap property ($m$-OGP) \cite{gamarnik2021overlap,gamarnik2025turing}. In order to define it, we need to introduce a distance between two arbitrary collisions. Take two arbitrary colliding pairs $\boldsymbol{c}_a, \boldsymbol{c}_b \in \{-1,1\}^{2N}$; we define their `\emph{external}' overlap as 
	\begin{equation}
		\label{eq:defExtOv}
		q_e(\boldsymbol{c}_a, \boldsymbol{c}_b) \equiv \frac{1}{2N}\left( \vecx_a^\top \vecx_b + \vecy_a^\top \vecy_b\right).
	\end{equation}
	The overlaps range from $-1$ to $1$. Let $m\geq 2$ be an integer. We say that the function $f_{\matA}(\vecx) : \mathbb{R}^N \to \mathbb{R}^P$ exhibits $m$-OGP for the collision-finding problem, with parameters $-1\leq q_1 \leq 1$ and $-1\leq\varepsilon_1<\varepsilon_2 \leq 1$ if, given a random $\matA$, it holds with high probability that there is no set of $m$ pairs $\boldsymbol{c}_a=(\vecx_a, \vecy_a)$, $a=1, \dots, m$ such that the following conditions are met:
	\begin{enumerate}
		\item $f_\matA(\vecx_a) = f_\matA(\vecy_a)$ with $\vecx_a \neq \vecy_a$, for every $a=1\ldots m$;
		\item $q_i(\boldsymbol{c}_a) = q_1$, 
		for every $a=1, \dots, m$; 
		\item \label{item2} $\varepsilon_1 \leq q_{e}(\boldsymbol{c}_a, \boldsymbol{c}_b)
		\leq \varepsilon_2$ for all $a \neq b$.
	\end{enumerate}
	The first condition imposes that each $\boldsymbol{c}_a$ is a collision. The parameter $q_1$ in the second condition controls the internal overlap of each collision as defined in~\eqref{eq::internal_overlap}; the third condition imposes a gap on the external overlap between each collision pair. We define, respectively, $q_1 = O(1) < 1$ and $q_1=1-o(1)$ as the `extensive' and `sub-extensive' regimes. 
	It is important to note that obstruction to stable algorithms does not require a specific value of $m$ \cite{gamarnik2021overlap}. In other words, independent of $m\geq 2$, $\alpha_{\mathrm{OGP}}^m(q_1)$ serves as an upper bound on the values of $\alpha$ where stable algorithms can find collisions with internal overlap $q_1$. Furthermore, if $m'>m$, then $m$-OGP implies $m'$-OGP, meaning that larger values of $m$ can provide stronger bounds on the region of algorithmic hardness. 
	
	The $m-$OGP has a simple geometric interpretation. It requires that for every $m$-tuple of collisions with a certain internal overlap, there are at least two collisions that are `close' or `far apart' in the sense of definition \eqref{eq:defExtOv}.\\
	
	In the following, we study the presence of $m-$OGP in the collision-finding problem for the function $f_\matA$, in the limit where both the dimensions of the input $N$ and the output $P$ of the function scale to infinity, with their ratio fixed to $\alpha = P/N = O(1)$. In this high-dimensional limit we show the existence of a region of $\alpha$ delimited by $\alpha_{\mathrm{OGP}}^m(q_1)$, such that for $\alpha>\alpha_{\mathrm{OGP}}^m(q_1)$ the $m$-OGP holds for the collision problem with internal overlap $q_1$. Moreover, $\alpha_{\mathrm{OGP}}^m(q_1)$ is strictly lower than the value $\alpha_c(q_1)$ above which no collisions with internal overlap $q_1$ can be found. We emphasize that the analytical formulas we derive in the Supplementary Material are valid for generic non-linearity $\varphi$. However, in the main text we focus specifically on the SWP activation function, due to its appealing cryptographic properties.
	
	\subsection{Existence of $m-$OGP in the Collision-Finding problem}
	\label{sec:m_OGP_analytics}
	
	The quantity of interest in establishing the presence of $m-$OGP is the number of {$m$-tuples} of collisions $\{\boldsymbol{c}_a=(\vecx_a,\vecy_a)\}_{a=1}^m$ with reciprocal external overlap $p$, and internal overlap $q_1$:
	\begin{multline}
		\mathcal{N}_m (p, q_1; \boldsymbol{A}) := \sum_{\{\boldsymbol{c}_a\}_{a=1}^m} \prod_{a = 1}^m \left[ \mathbb{X}_{\boldsymbol{A}}(\boldsymbol{c}_a)  \, \delta\!\left(q_{i}(\boldsymbol{c}_a) - q_1\right) \right] \\
		\times \prod_{a < b} \delta\left( q_e(\boldsymbol{c}^a, \boldsymbol{c}^b) - p\right).
	\end{multline}
	We say that the problem has an $m$-OGP when, for a given value of the internal overlap $q_1$, there exists a range of external overlaps $p$ where $\mathcal{N}_m (p, q_1; \boldsymbol{A})=0$ w.h.p.\ over the realization of the disorder. Note that this is analogous to condition \ref{item2} in the definition of $m$-OGP (see Sec.~\ref{sec:m_OGP}). The existence of such an excluded region can be proved by noticing that, since $\mathcal{N}_m(p,q_1; \matA)$ is a non-negative integer, Markov inequality gives
	\begin{equation}
		P(\mathcal{N}_m(p, q_1; \matA) > 0) \le  \mathbb{E}_{\matA} \mathcal{N}_m(p, q_1; \matA) = e^{m N \phi_m^{\mathrm{a}}(p, q_1)}.
		\label{eq:P_upperbound}
	\end{equation}
	where the annealed entropy $\phi_m^{\mathrm{a}}$ is defined as
	\begin{equation}
		\phi_m^{\mathrm{a}}(p, q_1) = \lim_{N \to \infty} \frac{1}{mN} \ln \mathbb{E}_{\boldsymbol{A}} \mathcal{N}_m(p, q_1; \boldsymbol{A}).
		\label{eq:f_m_definition}
	\end{equation}
	If, given $q_1$, for some value of $\alpha$ there exists an interval of $p$ where $\phi^{\mathrm{a}}_m(p,q_1) \le 0$, then \cref{eq:P_upperbound} proves that w.h.p.\ there are no collisions within such an external overlap range. We define $\alpha_{\mathrm{OGP}}^m(q_1)$ as the lowest value of $\alpha$ such that this condition is met. Notice that, because of Markov's inequality, $\alpha_{\mathrm{OGP}}^m(q_1)$ computed using the annealed entropy in~\eqref{eq:f_m_definition} is only an upper bound to the value $\alpha_{\mathrm{OGP}}$ above which an excluded region actually exists. 
	
	The computation of~\eqref{eq:f_m_definition} is detailed in the SM, here we only state the final result. One finds that $\phi_m^a(p, q_1)$ can be computed by maximizing a function $\phi_m(Q)$ over a suitable space of matrices $Q \in \bold{F}(p, q_1)$
	\begin{equation}
		\label{eq::saddle point}
		\phi_m^a(p, q_1) =  \max_{Q \in \mathcal{F}(p, q_1)} \phi_m(Q) \,.
	\end{equation}
	The detailed expression of $\phi_m(Q)$ is reported in the SM. $Q$ is a $2m \times 2m$ matrix that represents the covariance matrix of $m$-clones of colliding inputs. Namely denoting by $\alpha$ an index that runs over the $2m$ inputs of the $m$ collision clones, $Q$ is defined as
	\begin{equation}
		Q_{\alpha \beta} = \frac{1}{N} \sum_{k=1}^N w_{k \alpha} w_{k \beta} \,.
	\end{equation}
	The maximization over the entries of $Q$ in~\eqref{eq::saddle point} is performed over $\mathcal{F}(p, q_1)$, which is the set of covariance matrices $Q$ that fix the internal overlap $q_1$ of the $m$ collisions and the external overlap among them to $p$. In other words the entries of $Q$ should satisfy the following set of constraints
	\begin{subequations}
		\begin{align}
			\frac{1}{2}\left(Q_{2s-1, 2t-1} + Q_{2s, 2t} \right) &= p \,, \qquad \forall s<t \in [m] \\
			Q_{2s-1, 2s} &= q_1\,, \qquad \forall s \in [m]
		\end{align}
	\end{subequations}
	Finding the maximum in~\eqref{eq::saddle point} over the space of covariance matrices $Q$ can be done numerically; however, the complexity of such maximization increases considerably with $m$.
	We have therefore made an educated guess on the structure of the covariance matrix $Q$ achieving the global maximum of the function $\phi_m(Q)$. 
	We considered the so-called \emph{symmetric} ansatz, which imposes
	\begin{subequations}
		\label{eq::symmetric_ansatz}
		\begin{align}
			Q_{2s-1, 2t-1} = Q_{2s, 2t} &= p \,, \qquad \forall s<t \in [m] 
			\\ 
			Q_{2s-1, 2t} = Q_{2s, 2t-1} &= q_0 \,, \qquad \forall s<t \in [m]
		\end{align}
	\end{subequations}
	This ansatz considerably restricts the number of optimization parameters to just one (i.e. the parameter $q_0$). Moreover, this ansatz can be proved to be the right one in certain cases (see next section).  
	
	\begin{figure}[t]
		\centering
		\includegraphics[width=1.0\linewidth]{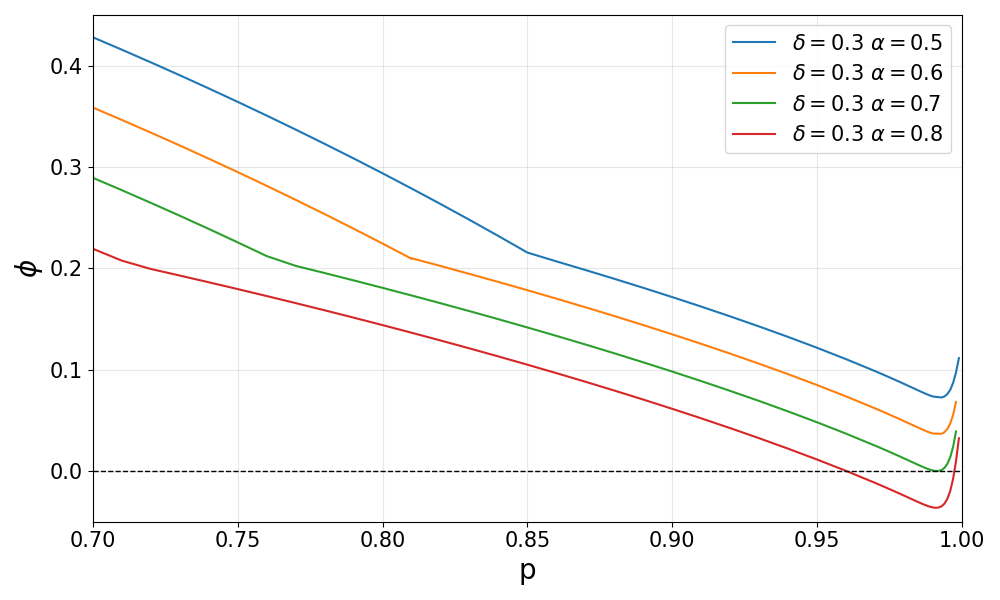}
		\caption{Typical behavior of $\phi^{\mathrm{a}}_m(p)$, for various values of $\alpha$, at $q_1=0.6$ and $m=5$. The activation is a square wave, with $\delta=0.3$. The onset value of the OGP is $\alpha=0.7$. For $\alpha>0.7$, an interval of external overlaps $p$ where $\phi<0$ is present.}
		\label{fig:phivsp}
	\end{figure}
	
	\begin{figure}[ht]
		\centering
		\includegraphics[width=1.0\linewidth]{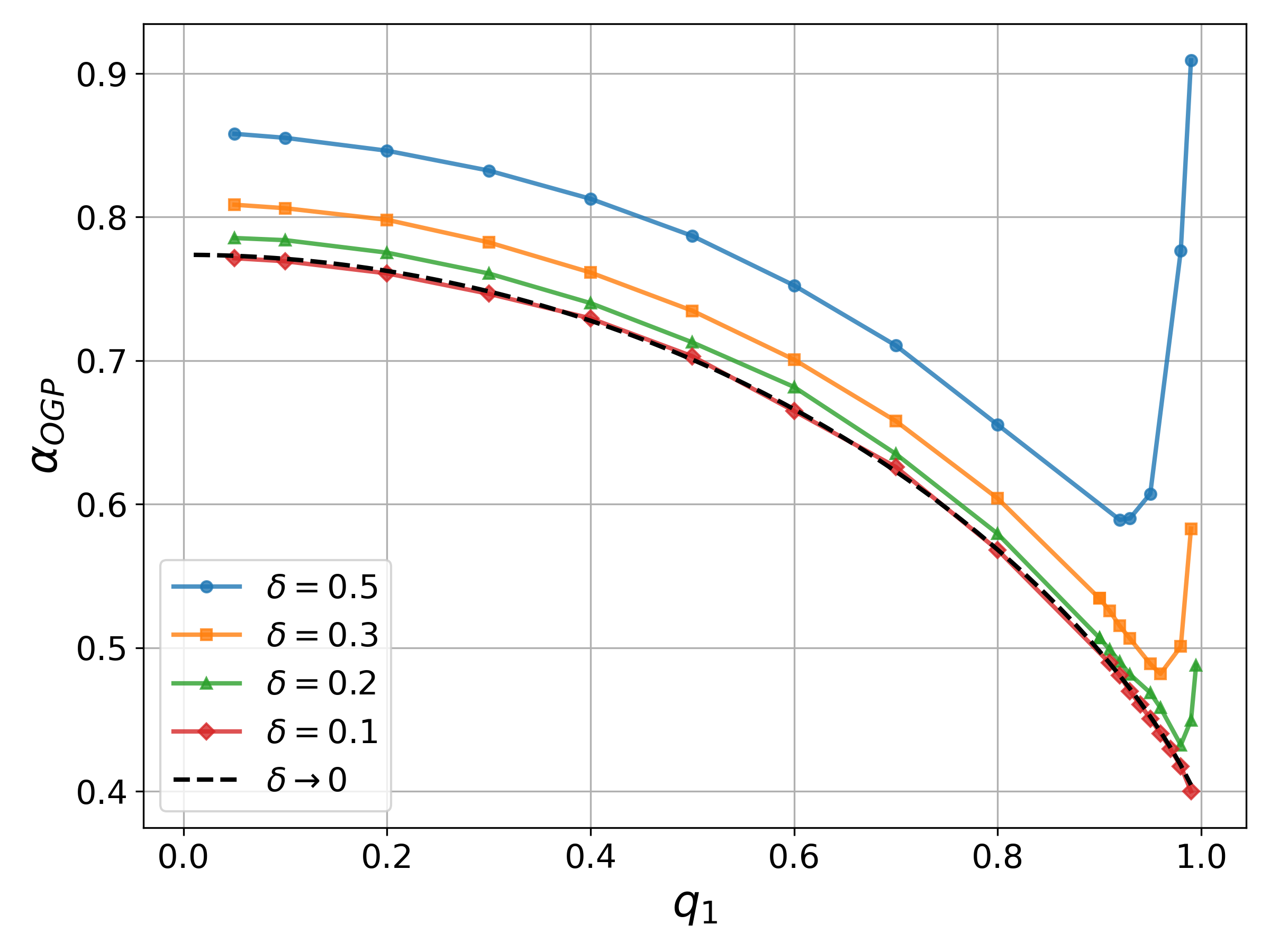}
		\caption{$\alpha_{OGP}^m$ versus the internal overlap $q_1$ for the Square Wave activation function, for various values of $\delta$, $m=5$.}
		\label{fig:alpha_OGP}
	\end{figure}
	
	\Cref{fig:phivsp} exemplifies the typical behavior of $\phi^{\mathrm{a}}_m(p)$, for various values of $\alpha$, including the onset value of the OGP at $\alpha\sim0.7$. Such curves are continuous, but not everywhere differentiable (see the SM for the details). 
	\Cref{fig:alpha_OGP} displays $\alpha_{\mathrm{OGP}}^m$ as a function of $q_1$ for $m=5$ and several values of $\delta$. Decreasing $\delta$ leads to a decrease of $\alpha_{\mathrm{OGP}}$ for all $q_1$. For any $\delta$ strictly larger than zero, the $\alpha_{\mathrm{OGP}}^m$ curve is non-monotonic in $q_1$, and start increasing when $q_1 \lesssim  1$, as observed for $\alpha_c(q_1)$. 
	We stress that this threshold is strictly below the one corresponding to the existence of collisions. Between $\alpha_{\mathrm{OGP}}^m(q_1)$ and $\alpha_c(q_1)$, despite the presence of exponentially many collisions we conjecture that finding one of them is algorithmically hard in the average case.

	\subsection{Strong Hashing Limit}
	\label{sec:strong_hashing_limit}
	
	The strong hashing limit (SHL) $\delta \to 0$ can be explicitly studied analytically. Moreover, it is a simple case in which the validity of the symmetric ansatz can be theoretically tested. 
	In particular, as we show in the SM, a detailed study of the case $m=2$ reveals that: i) the global maximization procedure~\eqref{eq::saddle point} leads to a symmetric~\eqref{eq::symmetric_ansatz} covariance matrix $Q$; ii) in the SHL, the optimum is reached at $q_0$ satisfying: 
	\begin{equation}
		\label{eq::vanishing_eigenvalue}
		q_0 = p + q_1 - 1.
	\end{equation}
	Condition \eqref{eq::vanishing_eigenvalue} implies that $Q$ has a vanishing eigenvalue. In the SM we show that, under the symmetric ansatz, the matrix $Q$ achieving the global maximum in the SHL satisfies the constraint~\eqref{eq::vanishing_eigenvalue} even for $m>2$, implying that $Q$ is not full rank in this case as well. 
	
\begin{figure}[ht]
		\centering
		\includegraphics[width=1.0\linewidth]{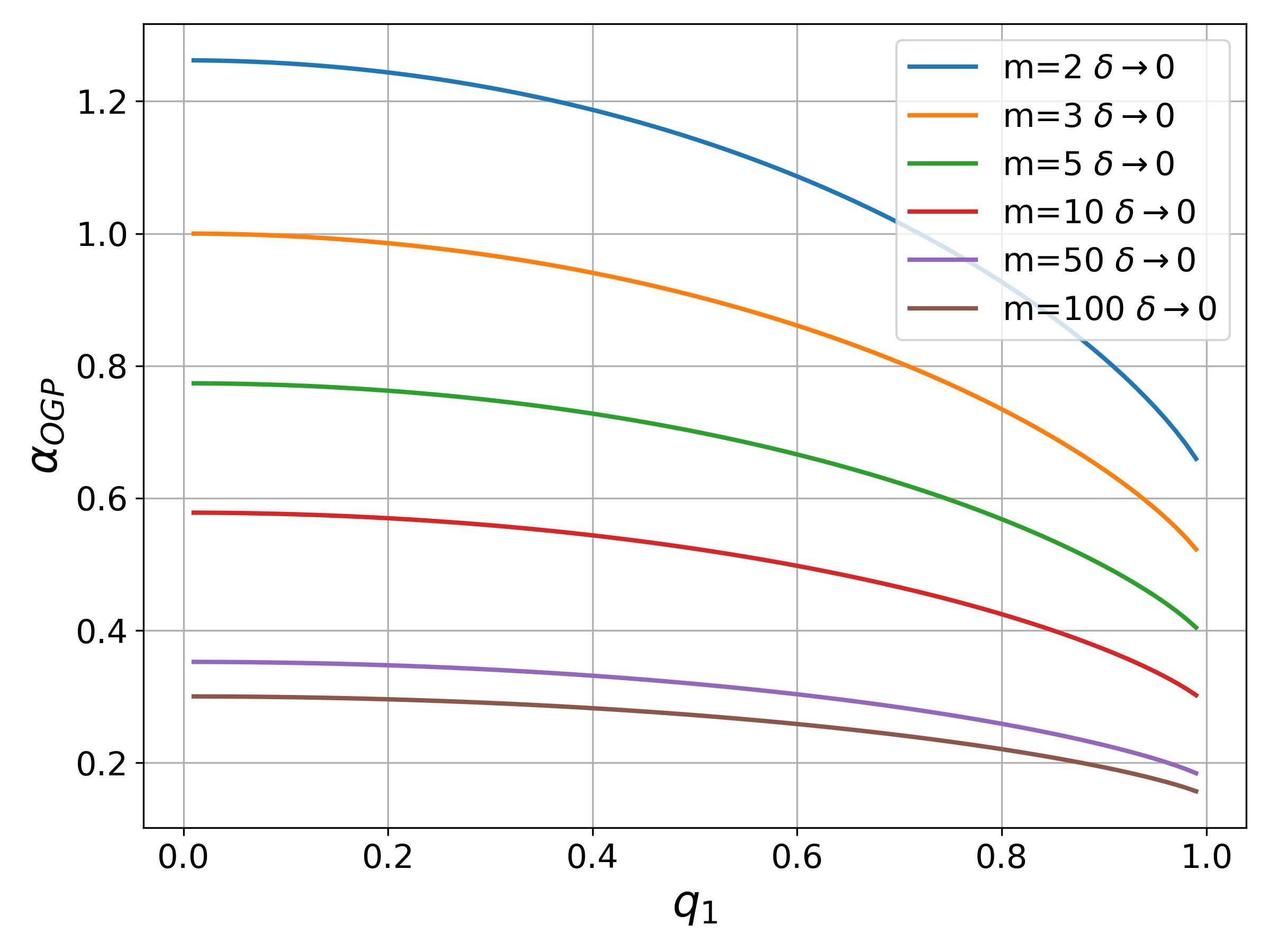}
		\caption{$\alpha_{\mathrm{OGP}}^m$ versus the internal overlap $q_1$ for the Square Wave activation function in the limit $\delta\to0$ for several values of $m$.}
		\label{fig:OGP_delta0}
	\end{figure}

	In the SHL, the annealed entropy curve becomes a monotonic function of the external overlap $p$. This allows us to explicitly find the value of $\alpha_{\mathrm{OGP}}^m(q_1)$ by performing the limit $p\to 1$ for every $m$. One finds:
	\begin{equation}
		\label{eq::OGP_strong_hashing}
		\alpha_{\mathrm{OGP}}^m(q_1) = \frac{\log 2 + H_B(q_1)}{\log(1+m)} \,.
	\end{equation}
	Note that this threshold is monotonically decreasing in $m$ and goes to $0$ for large $m$. In~\cref{fig:OGP_delta0} we show the OGP threshold given by~\eqref{eq::OGP_strong_hashing} as a function of $q_1$ and for different values of $m$. 
	
	In~\cref{fig:alpha_OGP} we plot the $\delta\to0$ OGP transition in black dashed for $m=5$ together with predictions for finite values of $\delta$. Notice that for $\delta=0.1$ already, the curve is practically indistinguishable from the SHL.
	
	Since we have established the validity of the symmetric ansatz only for $m=2$ in the SHL, this does not rule out the possibility that the true global maximum is not symmetric, especially for $\delta > 0$. We have therefore compared the analytical prediction of the free entropy under the symmetric ansatz, $\phi_m^{\mathrm{a}}(p,q_1)$, with a numerical estimate obtained by an exhaustive search in the space of $m$-tuples of collisions. The results presented in the SM are in agreement with the theory.
	
	\section{Hash functions from collision resistant Neural Networks}
	\label{sec::HFNN}
	\subsection{Collision Resistant Function}
	As already observed, when $\delta>0$ all $m$-OGP curves diverge in the limit $q_1\to 1$. Since an adversary is allowed to look for collisions with arbitrary value of $q_1$, the region $q_1\approx 1$ represents a potential flaw in the security of the hash function. To remedy this, recall that an \emph{error-correcting code} (ECC) is a function that maps distinct inputs into `codewords', separated by large Hamming distance. In order to obtain a collision resistant function, one can apply an ECC upstream of the network, restricting the space of inputs to $q_1\leq q_{\mathrm{code}}$. The results of \cref{sec:m_OGP_analytics} then imply that the composition $f_{\boldsymbol{A}} \circ ECC$ is collision resistant as long as $\alpha$ is higher than the OGP threshold for an adversarially chosen $q_1\leq q_{\mathrm{code}}$:
	\begin{equation}
		\alpha>\alpha_{\mathrm{adv}}(q_\mathrm{code}):=\max_{q_1\leq q_\mathrm{code}} \alpha_{\mathrm{OGP}}(q_1). 
	\end{equation}  
	
	\subsection{Collision Resistant Hash Function}
\begin{figure}[ht]
		\centering
		\includegraphics[width=1.0\linewidth]{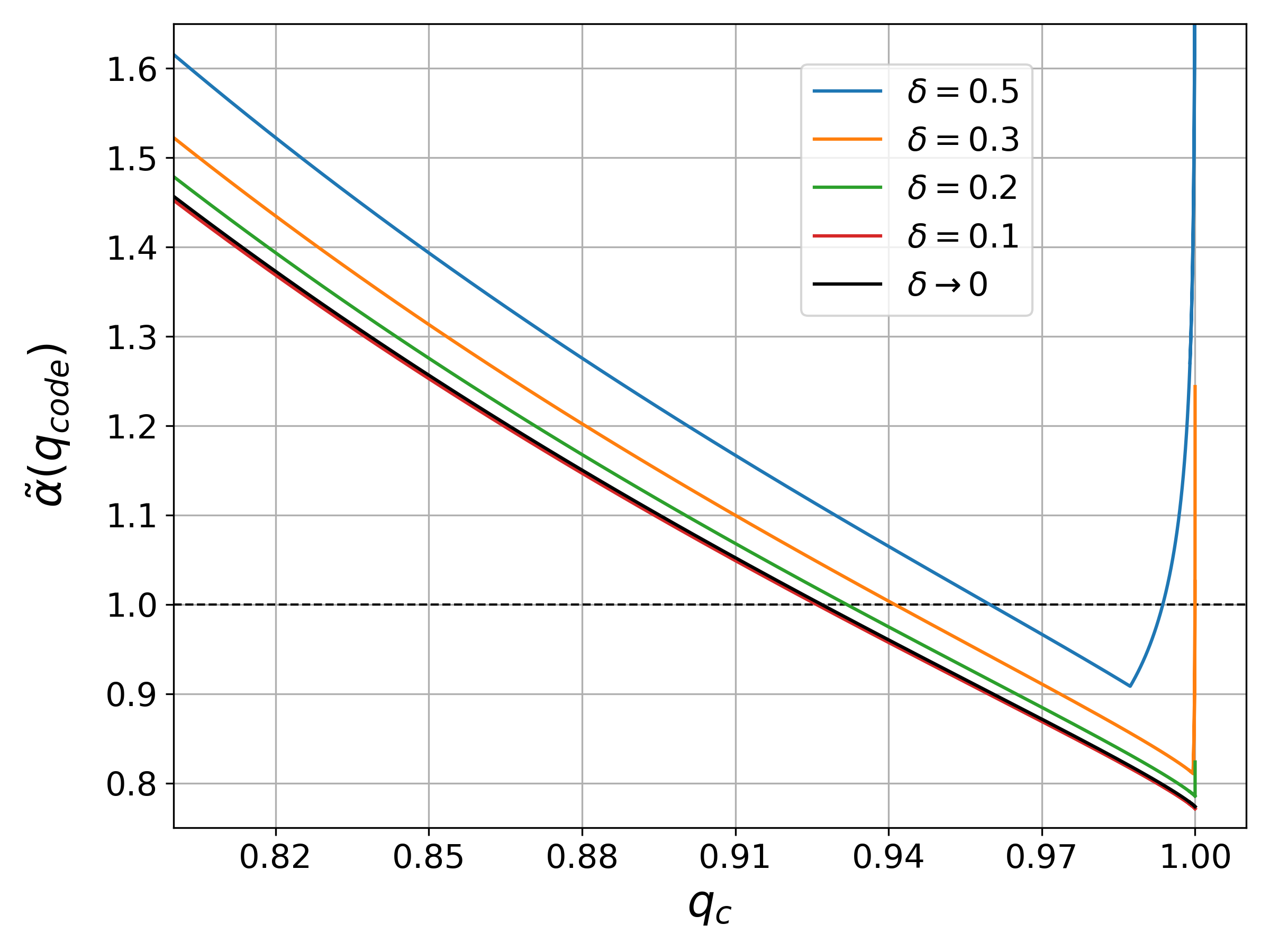 }
		\caption{The lowest compression rate of $NN\circ ECC$, i.e. $\alpha_{\mathrm{adv}}(q_{\mathrm{code}}) / r(q_{\mathrm{code}})$, as a function of $q_{\mathrm{code}}$, for different values of $\delta$. $\alpha_{\mathrm{adv}}(q_{\mathrm{code}})$ is given by a $m$-OGP computation with $m=5$.}
		\label{fig:compression_rate}
	\end{figure}

	We now show how, by composing the neural network with an ECC with appropriate $q_{\mathrm{code}}$ value, it is possible to build a CRH. We define the code rate $r$ as the ratio between the input and output bits of the ECC. The compression ratio of the neural network composed with the code is then $\tilde \alpha \equiv \alpha/r$. The Gilbert-Varshamov (GV) bound~\cite{mezard2009information} shows the existence of a binary code satisfying $r(q_\mathrm{code}) = 1 - H_2(q_\mathrm{code})$, where 
	\begin{equation*}
		{\textstyle H_2(q_\mathrm{code}) = -\frac{1-q_{\mathrm{code}}}{2} \log_2\frac{1-q_{\mathrm{code}}}{2} - \frac{q_{\mathrm{code}}+1}{2} \log_2\frac{q_{\mathrm{code}}+1}{2}.}
	\end{equation*}
	The lowest compression rate compatible with collision resistance is then 
	\begin{equation}
		\tilde \alpha(q_{\mathrm{code}})  =  \frac{\alpha_{\mathrm{adv}}(q_{\mathrm{code}})}{r(q_{\mathrm{code}})}
	\end{equation}
	obtained using a code matching the GV bound, and selecting the smallest $\alpha$ that guarantees hardness of collision-finding for $q_1\leq q_{\mathrm{code}}$.  \Cref{fig:compression_rate} shows $\tilde \alpha(q_{\mathrm{code}})$ as a function of $q_{\mathrm{code}}$, for $m=5$ and various $\delta$ values.  In the interval of $q_{\mathrm{code}}$ values where $\tilde \alpha(q_{\mathrm{code}}) < 1$, we conjecture that $f_{\boldsymbol{A}}\circ ECC$ is a CRH. \\

	\subsection{Comparison with Ajtai's Function.}
	\label{sec:Ajtai}
	
	Post-quantum cryptography seeks primitives whose security can be based on problems that are believed to remain intractable even for quantum algorithms. Among the various approaches, lattice-based cryptography has emerged as one of the most promising directions, offering efficient CRHs constructions whose security lies on worst-case to average case reductions.

	Ajtai's function~\cite{ajtai1996generating} stands as a landmark in lattice-based cryptography, as it laid the foundation for subsequent, more efficient lattice-based constructions. Formally, it acts on $\vecx \in \{0,1\}^N$ as:
	\begin{equation}
		f_\matA(\vecx) = \matA \vecx \quad \text{(mod $q$)}.
	\end{equation}
	Here, $q \in \bbN$ is a parameter and $\matA \in \bbZ_q^{P \times N}$ is chosen uniformly at random, where $\bbZ_q$ is a ring with $q$ elements. This function is shrinking when $P \log_2 q < N$. For certain values of $q, P=\poly(N)$ satisfying this bound, Ajtai's function can be shown to be a CRH, assuming it is hard to find short vectors in certain integer lattices \cite{goldreich2011collision,ajtai1996generating}. Namely, the security relies on the worst-case hardness of the \emph{short integer solution} (SIS) problem \cite{regev2009lattices} that for a matrix $\matA \in \bbZ_q^{n \times m}$ asks to find a `short'
	vector $\vecx \in \bbZ_q^m$ such that $\matA \vecx = 0$, a problem for which, for certain values of $q,M,N$, the only known algorithms run in time  $2^{\Omega(n)}$ \cite{pujol2009solving,schnorr1994lattice}, even with quantum computers. 
	
	Note that while our function bears some resemblance to Ajtai's function, as the oscillating nature of our activation is reminiscent of the mod operation. So our parameter $\delta$ plays a role similar to $q$ in Ajtai's function. However, there are also important differences. First of all, in Ajtai's case the worst to average case reduction, which is a crucial request for its collision resistance, is guaranteed when $q$ scales as $\mathrm{poly}(N)$, whereas in our case we can maintain $\delta=O(1)$. Secondly, the output spaces differ: our function maps to ${\pm 1}^P$, while Ajtai’s maps to $\mathbb{F}_q^P$. Moreover, in our case the entries of $\boldsymbol{A}$ are independent random variables which are drawn from an \emph{arbitrary} distribution. Note also that our security guarantee via the OGP criterion, only depends on the first two moments of such distribution. Finally, and most importantly, the periodicity of Ajtai's function is believed to be a central aspect of its hardness, while our results do not rely on the periodicity of the activation. Indeed, we have repeated the OGP analysis for an activation with a finite number $2K$ of oscillations, followed by a constant plateau, i.e.\ a `truncated' version of the SWP. Specifically, oscillations span a range $[-\gamma, \gamma]$, where $\gamma$ is a constant with $N$. As we show in the SM it is possible to find a region for $q_{\mathrm{code}}$ where our function composed with the ECC is a CRH. Preliminary checks indicate that qualitatively similar results also arise in a randomized version of the SWP, where the periodicity is broken by randomizing the points at which the activation function switches sign.
	
	These results suggest that what matters the most for our hardness criterion is not the periodicity of the function, but rather the frequency of sign switches of the activation, even if they are located in a limited region near the origin. 
	
	Qualitatively, our function can be considered a generalization of Ajtai's function, where one considers a specific bit of its output, when outputs in $\bbZ_q$ are represented as binary strings, rather than the least significant bit. The corresponding modification of Ajtai's function would not be secure due to the performance of the best-known lattice algorithms \cite{lenstra1982factoring,schnorr1988more}.
	

	\section{Algorithmic Attacks}
	\label{sec:AlgAtt}
	In this section, we present two algorithmic strategies for finding collisions: one involves a local search starting from a random point on the hypercube, where local moves are performed to explore potential collisions, while the other is based on approximate message passing.
	
	\subsection{Local Search From a Random Reference}
	
	Perhaps the most straightforward strategy to look for a collision in a binary perceptron is the following. Take a random binary configuration $\boldsymbol{w}$. To construct a collision, one can attempt a sequence of single bit-flips on the elements of $\boldsymbol{w}$, stopping as soon as the resulting vector $\boldsymbol{w}'$ collides with $\boldsymbol{w}$. In this section, we argue that this strategy is unfeasible if the number of patterns is proportional to $N$. In order to do so, let us count the average number $\mathcal{N}_t$ of collisions obtained by flipping $t=O(1)$ bits from the reference $\boldsymbol{w}$. One gets (see the SM):
	\begin{equation}
		\label{eq:Nt}
		\mathcal{N}_t\approx \binom{N}{t}\,e^{- a(\delta)\,t^{1/2}\,P\,N^{-1/2}},
	\end{equation}
	where $a(\delta)$ is a positive constant in $N$, which depends on $\delta$. \eqref{eq:Nt} implies that if $P=O(N)$, in the limit $N\rightarrow\infty$ there are no collisions in the finite-size neighborhood of typical vertices of the hypercube, indicating that a local search from a random reference $\boldsymbol{w}$ is an unfeasible strategy.   
	
	\subsection{Extensive Distances From a Random Reference}
	
	Consider again a random vertex of the hypercube $\boldsymbol{w}$. In this section, instead of looking at a finite-distance neighborhood of $\boldsymbol{w}$, we are interested in the algorithmic problem of finding collisions at an extensive distance. Note that in this case the probability of finding a collision is exponentially small in $N$, but there is an exponential number of configurations and the two can compensate (see \eqref{eq:Nt}). In order to do so, we study the performance of a message-passing algorithm inspired by statistical physics called reinforced approximate message passing (rAMP). Variants of the rAMP algorithms are known to be effective polynomial-time heuristics to solve inversion and second-preimage problems in perceptrons with binary synapses \cite{braunstein2006learning,baldassi2015subdominant,baldassi2015max,baldassi2016unreasonable,Baldassi2023}. We write in the SI the details of the implementation and the analysis of the performance with square wave activation. The complexity of rAMP is $O(T_{\mathrm{sol}}NP)$, where $NP\propto N^2$ comes from the fact that the algorithm involves matrix vector multiplications with the pattern matrix $\boldsymbol{A}$, and $T_{\mathrm{sol}}$ is the number of iterations required to find a solution. The quantity $T_{\mathrm{sol}}$ displays a power-law behavior $T_{\mathrm{sol}}\approx N^{b(\alpha)}$ in the system size, with a critical exponent $b(\alpha)\approx 1/(\alpha_{r}-\alpha)$. Therefore, the total complexity is $O\left(N^{2+b(\alpha)}\right)$. The exponent $b(\alpha)$ is usually extrapolated numerically (see \cite{baldassi2015max}). 
	
\begin{figure}[ht]
		\centering
		\includegraphics[width=1.0\linewidth]{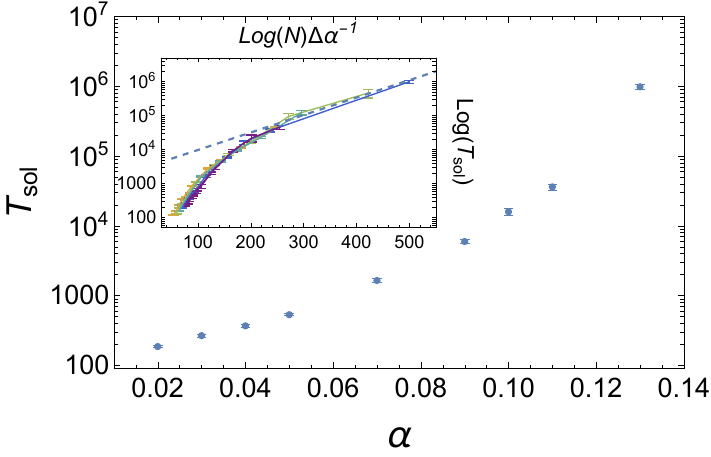}
		\caption{Number of iterations $T_{\mathrm{sol}}$ required by rAMP to find a solution for $N=8\times10^3$ and $\delta=0.6$, as a function of the constraint density $\alpha$. \emph{Inset}: $T_{\mathrm{sol}}$ has an exponential growth of the form $T_{\mathrm{sol}}\approx N^{b(\alpha)}$, with $b(\alpha)\propto (\alpha_r-\alpha)^{-1}\equiv \Delta\alpha^{-1}$ in the proximity of $\alpha_{r}\approx 0.15$. To test this behaviour, we show $\log{T_{\mathrm{sol}}}$ as a function of $\log{N}/\Delta\alpha$, for different values of $N$. The curves collapse for large $\log{N}/\Delta\alpha$. From bottom left to bottom right $N=n\times 10^3$, with $n=1,2,4,8,16,32$. The dashed line is a scaling function of the form $\exp{(a\, x+b)}$. }
		\label{fig:critical_exponent_alpha}
	\end{figure}

	In order to test the hardness of the collision-finding problem, we run rAMP on the SWP with $\delta=0.6$, where the annealed computation of the previous sections predicts the presence of an OGP. We find that rAMP outputs collisions having internal normalized overlap compatible with zero up to $\alpha_{\mathrm{rAMP}}\approx 0.15$, where $\alpha_{\mathrm{rAMP}}$ is estimated by a numerical characterization of the exponent $b(\alpha)$ (see Fig.~\ref{fig:critical_exponent_alpha}). We note that $\alpha_{\mathrm{rAMP}}$ is strictly smaller than the OGP estimate for $q_1=0$, that for $m=5$ and $\delta=0.6$ is $\alpha_{OGP}^{5}\approx 0.9$ (extrapolating from Fig.~\ref{fig:alpha_OGP}).
	
	\section{Conclusions}
	
	
	The square-wave perceptron, together with its truncated variant, defines a class of neural-network-based functions that, in a suitable regime, admit collisions while exhibiting features consistent with collision-resistance, as suggested by the presence of an overlap gap property. By combining these functions with error-correcting codes, we obtain a novel family of candidate collision-resistant hash functions. To our knowledge, this is the first instance where a geometric criterion rooted in statistical physics is employed to argue the security of cryptographic primitives. A direction for future work is to investigate whether more complex neural network architectures also display this property.

	
	$\\$
	
	\paragraph*{Acknowledgements}
	
	E.M.M. acknowledges the MUR-Prin 2022 funding Prot. 20229T9EAT, financed by the European Union (Next Generation EU). This paper is supported by PNRR-PE-AI FAIR project funded by the NextGeneration EU program. N.I.S. was supported by European Research Council (ERC) under the EU's Horizon 2020 research and innovation programme (Grant agreement No. 101019547). A.R. supported by European Research Council (ERC) under the EU's Horizon 2020 research and innovation programme (Grant agreement No. 101019547) and Cariplo CRYPTONOMEX grant.

	\bibliography{references}
	
	\clearpage
	
	\title{Are Neural Networks Collision Resistant? \\ SUPPLEMENTAL MATERIAL}
	
	\maketitle
	
	\onecolumngrid 
	
	\tableofcontents{}
	
	\appendix

\section{The entropy of the $m-$cloned system}
\label{app:entropy_clones}

\subsection{Recap of the main definitions}
As discussed in the main text, we are interested in computing the following free entropy:
\begin{equation}
	\phi_m^{\mathrm{a}}(p, q_1) = \lim_{N \to \infty} \frac{1}{mN} \log \mathbb{E}_{\boldsymbol{A}} \mathcal{N}_m(p, q_1; \boldsymbol{A})
	\label{eq:f_m_definition_app}
\end{equation}
where
\begin{equation}
	\mathcal{N}_m (p, q_1; \boldsymbol{A}) \equiv \sum_{\{\boldsymbol{c}_a\}_{a=1}^m} \prod_{a = 1}^m \, \mathbb{X}_{\boldsymbol{A}}(\boldsymbol{c}_a) \prod_{a < b} \delta\left(q_e(\boldsymbol{c}_a, \boldsymbol{c}_b) - p\right) \prod_{a=1}^m \delta\left(q_i(\boldsymbol{c}_a) - q_1 \right)
\end{equation}
represents the number of $m$-uples of collisions $\boldsymbol{c}_a \equiv (\vecx_a, \vecy_a)$, $a \in [m]$, that are at a mutual overlap $p \in [0, 1]$ and with an \emph{internal overlap} $q_1\in [-1, 1]$. We consider as internal overlap of a collision simply the scalar product:
\begin{equation}
	\label{eq::internal_overlap_collision}
	q_i(\boldsymbol{c}_a) = \frac{1}{N} \vecx_a \cdot \vecy_a.
\end{equation}
Note that the function $\mathbb{X}_{\boldsymbol{A}}(\boldsymbol{c})$ 
\begin{equation}
	\label{eq::constraint_collision}
	\mathbb{X}_{\matA}(\boldsymbol{c}) = \prod_{\mu=1}^P\Theta\left[f_{\matA}(\vecx)_{\mu} f_{\matA}(\vecy)_{\mu} \right]= \prod_{\mu=1}^P  \Theta \left[\varphi\left(\frac{1}{\sqrt{N}} \vecx \cdot\boldsymbol A^{\mu} \right)  \varphi\left(\frac{1}{\sqrt{N}} \vecy \cdot\boldsymbol A^{\mu} \right) \right]
\end{equation}
imposes that $\boldsymbol{c} = (\vecx, \vecy)$ is a collision. With $\boldsymbol{A}^\mu$ we denoted the $\mu-$th row of the data matrix $\boldsymbol{A}\in \mathbb{R}^{P \times N}$, and with $f_\matA(\bullet)_{\mu}$ the $\mu$-th element of vector $f_\matA(\bullet)$. 

As stated in the main text, the \emph{annealed} entropy in~\eqref{eq:f_m_definition_app} gives an upper bound to the existence of such constrained $m$-uplets of collisions, thanks to Markov's inequality.

\subsection{Overlap among collisions}

We consider an overlap function between two collisions $\boldsymbol{c}_a= (\vecx_a, \vecy_a)$ and $\boldsymbol{c}_b= (\vecx_b, \vecy_b)$ of the form
\begin{equation}
	\label{eq::external_overlap_collision}
	q_e(\boldsymbol{c}_a, \boldsymbol{c}_b) = \frac{1}{2N} \left( \vecx_a^\top \vecx_b + \vecy_a^\top \vecy_b \right).
\end{equation}
For simplicity, we will name this constraint as the \emph{external overlap} between two collisions. 

\subsection{Computation of $\phi_m^{\mathrm{a}}(p, q_1)$} \label{app::m-clones_computation}

The computation of $\mathbb{E}_{\boldsymbol{A}} \mathcal{N}_m(p, q_1; \boldsymbol{A})$ follows from simple statistical physics methods. With a slight abuse of notation we will denote from now on the elements forming a collision with the same letter $\boldsymbol{w}$ as $\boldsymbol{c}_a = (\boldsymbol{w}_a^{(1)}, \boldsymbol{w}_a^{(2)})$. We will also use the index $a \in [m]$ as the one running over the number of clones $m$, whereas with $s = 1, 2$ as the one that runs over the two weight components composing each collision. 
We can use the identity
\begin{equation}
	1=\prod_{\mu,a,s}\int\, dh^{\mu a}_s\,\delta\left(h^{\mu a}_s-\frac{1}{\sqrt{N}}\matA^{\mu}\cdot \boldsymbol{w}^{(s)}_a\right) = \prod_{\mu,a,s}\int\, \frac{dh^{\mu a}_s d \hat h^{\mu a}_s}{2\pi} \, e^{i \hat h_s^{\mu a} \left( h_s^{\mu a} - \frac{1}{\sqrt{N}}\matA^{\mu}\cdot \boldsymbol{w}^{(s)}_a \right) }
\end{equation}
to extract the dependence on the matrix $\matA$ in $\prod_a \mathbb{X}_{\boldsymbol{A}}(\boldsymbol{c}_a)$. The average over $\matA$ is then straightforward
\begin{equation}
	\prod_{k=1}^N \prod_{\mu = 1}^P \mathbb{E}_{A_k^\mu} e^{- i\frac{A_k^\mu}{\sqrt{N}} \left( \sum_{a s} \hat h^{\mu a}_s w_{ak}^{(s)} \right)} = \prod_{k=1}^N \prod_{\mu = 1}^P e^{-\frac{1}{2N} \left( \sum_{\alpha} \hat h^{\mu}_\alpha w_{k \alpha} \right)^2} = \prod_{\mu = 1}^P e^{-\frac{1}{2} \sum_{\alpha \beta} Q_{\alpha \beta} \hat h^{\mu}_\alpha \hat h^{\mu}_\beta} 
\end{equation}
where we flattened the indices $(a, s)$ into a collision index $\alpha \in [2m]$. 
We have also introduced the $2m\times 2m$-dimensional overlap matrix $Q_{\alpha \beta}$ which is defined as
\begin{equation}
	\label{eq::overlap}
	Q_{\alpha \beta} \equiv \frac{1}{N} \sum_{k=1}^N w_{k \alpha} w_{k \beta}
\end{equation}
which can be equivalently expressed in terms of indices $\alpha, \beta \in [2m]$ as a matrix that we will call $Q_{\alpha \beta}$. The corresponding $2y \times 2y$-dimensional matrix will be called as $Q$. Enforcing the definition in~\eqref{eq::overlap} via a delta function
\begin{equation}
	\label{eq::constraints_for_Q}
	1=
	\prod_{\alpha < \beta} \int d Q_{\alpha \beta} \, \delta\left(Q_{\alpha \beta} - \frac{1}{N} \boldsymbol{w}_\alpha \cdot \boldsymbol{w}_\beta \right)
\end{equation}
one finally gets that the index $k\in [N]$ and $\mu \in [P] = [\alpha N]$ become decoupled. 
Notice that, the matrix $Q_{\alpha \beta}$ is symmetric and $Q_{\alpha \alpha} = 1$ for any $\alpha \in [2m]$ as the weights are binary $w_{k\alpha} = \pm 1$. This is the reason why we need to impose $2y(2y-1)$ constraints in equation~\eqref{eq::constraints_for_Q}.  
The expected value of $\mathcal{N}_m(p, q_1; \boldsymbol{\matA})$ can be therefore written as
\begin{equation}
	\begin{split}
		\mathbb{E}_{\boldsymbol{\matA}} \left[\mathcal{N}_m(p, q_1; \boldsymbol{\matA})\right] &= 
		\int_{Q \in \mathcal{F}(p, q_1)} \prod_{\alpha < \beta} d Q_{\alpha \beta} \, e^{N m \phi_m (Q)}
		\label{eq:App_f_1}
	\end{split}
\end{equation}
with
\begin{subequations}
	\begin{align}
		\phi_m(Q) &\equiv G_S(Q) + \alpha G_E(Q) \label{eq:Si_form1}\\
		G_S(Q) &\equiv  \lim_{N\to \infty} \frac{1}{N} \log \left[ \sum_{\boldsymbol{w}_\alpha} \prod_{\alpha < \beta} \delta \left( Q_{\alpha \beta} - \frac{1}{N} \boldsymbol{w}_\alpha \cdot \boldsymbol{w}_\beta \right) \right]
		\label{eq:Gs_form1}\\
		G_E(Q) &\equiv 
		\frac{1}{m} \ln \int \mathcal{D}_Q (u_1, v_1, \dots u^m, v^m) \, \prod_{a=1}^m \Theta\left[ \varphi(u^a)\varphi(v^a) \right] = \frac{1}{m} \ln \int \mathcal{D}_Q \boldsymbol{h} \, \prod_{a=1}^m \Theta\left[ \varphi(h^{2a-1})\varphi(h^{2a}) \right],
		\label{eq:Ge_form1}
	\end{align}
\end{subequations}
where $\mathcal{D}_Q$ is a multivariate Gaussian measure over the variables $\boldsymbol{h} = (u_1, v_1, \dots u^m, v^m)$ with zero mean ad covariance given by the $2m \times 2m$-dimensional matrix of overlaps $Q$. $\mathcal{F}(p, q_1)$ represents the set of $2m\times 2m$ overlap matrices $Q$ whose elements satisfy the constraint that the external~\eqref{eq::external_overlap_collision} and internal overlaps~\eqref{eq::internal_overlap_collision} of the collisions are respectively $p$ and $q_1$. 
The integral in \Cref{eq:App_f_1} can be computed at leading order in the limit $N\to\infty$ through the Saddle Point method, i.e.
\begin{equation}
	\mathbb{E}_{\boldsymbol{\matA}} \left[\mathcal{N}_m(p, q_1; \boldsymbol{\matA})\right]= e^{Nm\phi_m^\star(p, q_1)} \,,
\end{equation}
where
\begin{equation}
	\label{eq::max_procedure}
	\phi_m^\star(p, q_1) \equiv \max_{Q \in \mathcal{F}(p, q_1)} \mathcal{G}_m(Q).
\end{equation}

\subsection{The admissible set of covariance matrices}	
We say that an overlap matrix $Q \in \mathcal{F}(p, q_1)$ is \emph{admissible} if there exists a sequence of binary inputs $\{w_{k \alpha}=\pm 1\}$, $k\in [N]$, $\alpha\in[2m]$ such that 
\begin{equation}
	\label{eq::admissible_matrix}
	Q_{\alpha \beta} = \frac{1}{N} \sum_{k=1}^N w_{k \alpha} w_{k \beta} \,.
\end{equation}
Admissible matrix are symmetric, have all elements in the interval $[-1, 1]$ and have the diagonal elements equal to 1. Moreover another necessary condition for $Q$ to be admissible is that it is positive semi-definite, indeed for any vector $u$: 
\begin{equation}
	\label{eq::Q_positive_definite}
	\sum_{\alpha \beta} u_\alpha Q_{\alpha \beta} u_\beta = \frac{1}{N} \sum_{k=1}^N \left(\sum_\alpha w_{k\alpha} u_\alpha\right)^2\ge 0 \,.
\end{equation}
In addition being $Q \in \mathcal{F}(p, q_1)$ the matrix element should satisfy the following constraints
\begin{subequations}
	\begin{align}
		\frac{1}{2}\left(Q_{2s-1, 2t-1} + Q_{2s, 2t} \right) &= p \,, \qquad \forall s<t \in [m] \\
		Q_{2s-1, 2s} &= q_1\,, \qquad \forall s \in [m]
	\end{align}
\end{subequations}
With a slight abuse of language, we will simply say that the set $\mathcal{F}(p, q_1)$ denotes the set of admissible covariance matrices. 

Another important aspect of admissible covariance matrices concerns the possibility of some eigenvalues being zero. If the matrix $Q$ has an eigenvector $u$ with eigenvalue 0,~\eqref{eq::Q_positive_definite} implies that there must exist a set of $2m$ spins $\tau_\alpha = \pm 1$ such that
\begin{equation}
	\label{eq::Q_zero_eigenvalue_contraint}
	\sum_{\alpha =1}^{2m} \tau_{\alpha} u_\alpha = 0 \,.
\end{equation}

\subsection{The entropic term}

By introducing the Fourier integral representation of the delta function,  the entropic term can be written as
\begin{equation}
	G_S(Q) = \min_{\hat Q} \left[- \frac{1}{2m} \mathrm{Tr}[Q \hat Q]+ \frac{1}{m}\log 
	\sum\nolimits_{ \boldsymbol{w}} e^{\frac{1}{2} \boldsymbol{w}^T \hat Q \boldsymbol{w}} \right],
\end{equation}
where we have introduced a conjugated variable matrix $\hat Q$ which enforces each of the constraints $Q_{\alpha \beta} = \frac{1}{N} \sum_{k=1}^N w_{i \alpha} w_{i\beta}$ with $\alpha < \beta$. Note that due to the binary nature of the inputs, we do not need to impose any constraint for $Q_{\alpha \alpha} = 1$ $\alpha \in [2m]$; so we consider $\hat Q_{\alpha \alpha} = 0$ when needed. 

From this representation we can find a nice property of the $G_S(Q)$ function, namely that it is a \emph{concave} function. Indeed, note that the function to be minimized in $G_S$ is convex. The concave property therefore follows from the fact that $G_S(Q)$ is the Legendre transform of this convex function.

\subsection{Energetic term for the Square Wave activation function}

We restrict here the discussion to the main activation function which we have analyzed in this paper, that is the square wave, which is
\begin{equation}
	\varphi(h) \equiv \text{sgn}\left(\frac{h}{2\delta} \,\mathrm{mod}\,1\right) = \mathrm{sign}\left[\sin \left( \frac{\pi}{\delta} h\right)\right],
\end{equation}
where $\delta$ tunes the width of the oscillations. 
Since this activation is periodic with period $2\delta$ we can conveniently evaluate the energetic term via Fourier transform. 

Let $\mathcal{P}([m])$ be the set of subset of the set $[m]$. For each element $\gamma \in \mathcal{P}([m])$ we will denote by $|\gamma|$ its cardinality. The integrand in the energetic term can be written for the square wave activation as
\begin{equation}
	\prod_{a=1}^m \Theta\left[ \varphi(h^{2a-1})\varphi(h^{2a}) \right] = \frac{1}{2^m} \prod_{a=1}^m \left[1 + \varphi(h^{2a-1})\varphi(h^{2a}) \right] = \frac{1}{2^m} + \frac{1}{2^m} \sum_{r=1}^m \sum_{\substack{\gamma \in \mathcal{P}([m]) \\ |\gamma| = r}} \prod_{g \in \gamma} \varphi(h^{2g-1})\varphi(h^{2g}).
\end{equation}
In the following, we will denote by $\boldsymbol{h}_\gamma$ the $2r$ dimensional vector containing the variables $h^{2g-1}$ and $h^{2g}$, $\forall g \in \gamma$. 
We define the function of $2r$ variables as
\begin{equation}
	f(\boldsymbol{h}_\gamma) = \prod_{g \in \gamma} \varphi(h^{2g-1})\varphi(h^{2g}),
\end{equation}
so that the energetic term can be expressed as
\begin{equation}
	G_E = \frac{1}{m} \ln \left[ \frac{1}{2^m} + \frac{1}{2^m}   \sum_{r=1}^m \sum_{\substack{\gamma \in \mathcal{P}([m]) \\ |\gamma| = r}} \int \mathcal{D}_{Q_\gamma} \boldsymbol{h}_\gamma \, f(\boldsymbol{h}_\gamma) \right].
\end{equation}
In the previous expression, we have integrated over the variables $h$ not appearing in the integrand that depends on $\boldsymbol{h}_\gamma$. We also defined
$Q_\gamma$ as the $2r \times 2r$ matrix obtained from $Q$ by removing the $(2g-1)$-th and $2g$-th rows, as well as the corresponding columns, for all $g \notin \gamma$.

The $2r$ dimensional multivariate Gaussian integration can be computed analytically by using the Fourier transform of the function $f(\boldsymbol{h}_\gamma)$
\begin{equation}
	\int \mathcal{D}_{Q_\gamma} \boldsymbol{h}_{\gamma} \, f(\boldsymbol{h}_{\gamma}) = \sum_{\boldsymbol{n} \in \mathbb{Z}^{2r} } \hat f(\boldsymbol{n}) \int \mathcal{D}_{Q_\gamma} \boldsymbol{h} \, e^{- i \frac{\pi}{\delta} \boldsymbol{n}^T \boldsymbol{h}_\gamma} = \sum_{\boldsymbol{n} \in \mathbb{Z}^{2r} } \hat f(\boldsymbol{n}) \, e^{- \frac{\pi^2}{2\delta^2} \boldsymbol{n} Q_\gamma \boldsymbol{n}} \,,
\end{equation}
where $\hat f$ is the Fourier transform of $f$ which reads
\begin{subequations}
	\begin{align}
		\hat f(\boldsymbol{n}) &= \prod_{i=1}^{2r} \hat \varphi(n_i)\\
		\hat \varphi(n) &= \frac{1}{\delta} \int_{-\delta}^{\delta} dh \, \varphi(h) \sin \left(\frac{\pi}{\delta} n h\right) = \frac{4 \cos(n \pi) \sin^2\left(\frac{n\pi}{2}\right)}{n \pi},
	\end{align}
\end{subequations}
which is equal to $-4/(\pi n)$ if $n$ is odd and $0$ otherwise. We therefore get the following representation of the energetic term for the square wave activation
\begin{equation}
	\label{eq:Ge_fourier}
	G_E = \frac{1}{m} \ln \left[ \frac{1}{2^m} + \frac{1}{2^m}   \sum_{r=1}^m \sum_{\substack{\gamma \in \mathcal{P}([m]) \\ |\gamma| = r}} \left(-\frac{4}{\pi^2}\right)^r \sum_{\boldsymbol{n} \in \mathbb{Z}^{2r} } \frac{1}{\prod_{i=1}^{2r} \left(2n_i+1\right)} \, e^{- \frac{\pi^2}{2\delta^2} (2\boldsymbol{n}+1)^T Q_\gamma (2\boldsymbol{n}+1)}  \right].
\end{equation}

\subsection{Symmetric Ansatz and Hubbard-Stratonovich representation} \label{app::Symmetric_Ansatz}
We now proceed under the assumption that such optimum can be found among the class of symmetric matrices
\begin{subequations}
	\begin{align}
		Q_{2s-1, 2t-1} = Q_{2s, 2t} &= p \,, \qquad \forall s<t \in [m] 
		\\ 
		Q_{2s-1, 2t} = Q_{2s, 2t-1} &= q_0 \,, \qquad \forall s<t \in [m]
	\end{align}
\end{subequations}
Similarly, we will consider a symmetric ansatz over the conjugated matrix
\begin{subequations}
	\begin{align}
		\hat Q_{2s-1, 2t-1} = \hat Q_{2s, 2t} &= \hat p \,, \qquad \forall s<t \in [m] 
		\\ 
		\hat Q_{2s-1, 2t} = \hat Q_{2s, 2t-1} &= \hat q_0 \,, \qquad \forall s<t \in [m] \\
		\hat Q_{2s-1, 2s} &= \hat q_1 \,, \qquad \forall s \in [m].
	\end{align}
\end{subequations}
For example, for $m=3$, one has:
\begin{equation}
	Q = 
	\left(
	\begin{array}{cc|cc|cc}
		1 & q_1 & p & q_0 & p & q_0 \\
		q_1 & 1 & q_0 & p & q_0 & p \\\hline
		p & p_0 & 1 & q_1 & p & q_0 \\
		q_0 & p & q_1 & 1 & q_0 & p \\\hline
		p & p_0 & p & q_0 & 1 & q_1 \\
		q_0 & p & q_0 & p & q_1 & 1 \\
	\end{array}
	\right ),
	\,\quad 
	\hat{Q} = 
	\left(
	\begin{array}{cc|cc|cc}
		0 & \hat{q}_1 & \hat{p} & \hat{q}_0 & \hat{p} & \hat{q}_0 \\ 
		\hat{q}_1 & 0 & \hat{q}_0 & \hat{p} & \hat{q}_0 & \hat{p} \\ \hline
		\hat{p} & \hat{p}_0 & 0 & \hat{q}_1 & \hat{p} & \hat{q}_0 \\
		\hat{q}_0 & \hat{p} & \hat{q}_1 & 0 & \hat{q}_0 & \hat{p} \\ \hline
		\hat{p} & \hat{p}_0 & \hat{p} & \hat{q}_0 & 0 & \hat{q}_1 \\
		\hat{q}_0 & \hat{p} & \hat{q}_0 & \hat{p} & \hat{q}_1 & 0 \\
	\end{array}
	\right ).
\end{equation}
This reasonable assumption can be verified for small $m$ numerically by solving the optimization problem for general matrices and checking that symmetry holds at the optimum. Furthermore, in the SHL and $m=2$ it can be proved rigorously (see~\cref{app::2OGP_RHF}). 
Within this ansatz, one can perform a Hubbard-Stratonovich transformation to decouple terms in $G_S$ and $G_E$ that depend on different clone indexes. As a result, the parameter $m$ appears as an exponent, rather as the number of integration variables in the expression, and can be promoted to a real value as in standard replica computations. For example, the entropic term reads
\begin{equation}
	\begin{split}
		G_S &= \nonumber - q_1 \hat q_1 - (m-1)\, p\, \hat p - (m-1)\, q_0\, \hat q_0 \\
		&\nonumber - (\hat p + \hat q_1 - \hat q_0) + \frac{1}{m} \log \int D u_1 Du_2 \left[2 \cosh\left(\sqrt{2(\hat p - \hat q_0)} \,u_1\right) + 2 e^{2(\hat q_1 - \hat q_0)} \cosh\left(\sqrt{2(\hat p + \hat q_0)}\, u_2\right) \right]^m \\
		&= - q_1 \hat q_1 - (m-1) p \hat p - (m-1) q_0 \hat q_0 \\
		& \nonumber - (\hat p + \hat q_1 - \hat q_0) + \frac{1}{m} \log \sum_{k=0}^m \binom{m}{k} e^{2 (m-k) (\hat q_1 - \hat q_0)} \left[ \sum_{s=0}^k \binom{k}{s} e^{(k-2s)^2 (\hat p - \hat q_0)} \right] \left[ \sum_{s=0}^{m-k} \binom{m-k}{s} e^{(m-k-2s)^2 (\hat p + \hat q_0)} \right]
		\label{eq::Gs_useful}
	\end{split}
\end{equation}
where $Dz \equiv dz \, \frac{e^{-z^2/2}}{\sqrt{2\pi}}$. The energetic term can be written instead as
\begin{equation}
	\begin{split}
		G_E &\equiv \frac{1}{m}\log \int \prod_{a s} \frac{d h^{a}_s d \hat h^{a}_s}{2\pi} e^{i h^{a}_s \hat{h}^{a}_s} \prod_{a=1}^m \Theta\left[ \prod_{s=1}^2 \varphi(h^a_s) \right] e^{-\frac{1}{2} \sum_{ab, st} q^{ab}_{st} \hat h^a_s \hat h^b_t }\\
		&=\frac{1}{m}\log \int \prod_{a s} \frac{d h^{a}_s d \hat h^{a}_s}{2\pi} e^{i h^{a}_s \hat{h}^{a}_s} \prod_{a=1}^m \Theta\left[ \prod_{s=1}^2 \varphi(h^a_s) \right] \\
		&\times e^{-\frac{1-q_1 - p + q_0}{2} \sum_{as} (\hat h^a_s)^2  -\frac{q_1 -q_0}{2} \sum_{a} \left(\sum_s \hat h^a_s\right)^2 - \frac{p-q_0}{2} \sum_{s} \left( \sum_a \hat h^a_t\right)^2 - \frac{q_0}{2} \left(\sum_{as} \hat h^a_s \right)^2} \\
		&= \frac{1}{m}\log \int \prod_s Du_s Dx \left[ \int Dz \prod_{s} D h_s \, \Theta\left( \prod_s \varphi(\sqrt{1-q_1 - p + q_0}\, h_s + \sqrt{p-q_0}\, u_s + \sqrt{q_0}\, x + \sqrt{q_1 - q_0}\,z)\right) \right]^m.
		\label{eq::Ge_useful}
	\end{split}
\end{equation}
Using the property of Heaviside function $\Theta\left( \prod_{s=1}^2 \varphi(a_s)\right) = \prod_{s=1}^2 \Theta\left( \varphi(a_s)\right) + \prod_{s=1}^2 \Theta\left( - \varphi(a_s)\right)$ and performing some 2D rotations of variables one gets a simplified expression of the energetic term
\begin{equation}
	G_E = \frac{1}{m}\log \int \prod_s Du_s \left[ \int Dz  \, F_\varphi\!\left( a(u_1, u_2, z), a(u_1, -u_2, z); \sigma
	\right) \right]^m,
	\label{eq:Ge_general} 
\end{equation}
where we have defined the quantities
\begin{subequations}
	\label{eq::convenient_quantities}
	\begin{align}
		a(u_1, u_2, z) &\equiv \sqrt{\frac{p+q_0}{2}}u_1-\sqrt{\frac{p-q_0}{2}}u_2+\sqrt{q_1-q_0}z \\
		\sigma &\equiv \sqrt{1-q_1 - p + q_0} 
	\end{align}
\end{subequations}
and the $\varphi$-dependent functions
\begin{subequations}
	\begin{align}
		F_\varphi(x, y; \sigma) &\equiv 1 - I_\varphi(x; \sigma) - I_\varphi(y; \sigma) + 2 I_\varphi(x; \sigma) I_\varphi(y; \sigma) \\
		I_\varphi(x; \sigma) &\equiv \int Dh \, \Theta\left[ \varphi\left(\sigma h + x \right)\right] \,.
		\label{eq::Kernel_Generic}
	\end{align}
\end{subequations}
With such simplifications, $\phi_m(Q, \hat Q)$ can be optimized numerically, for each value of the external parameters $m$, $\alpha$, $p$. \\

\subsubsection{Explicit Formulas for some activation functions}
We list here the expression of the Kernel $I_\varphi(x; \sigma)$ defined in~\eqref{eq::Kernel_Generic} for some case of interest.

\begin{itemize}
	\item Asymmetric Binary Perceptron (ABP): $\varphi(h) = h$. One gets
	\begin{equation}
		I_{\mathrm{ABP}}(x; \sigma) = H\left(- \frac{x}{\sigma} \right)
	\end{equation}
	\item Symmetric Binary Perceptron (SBP): $\varphi_\kappa(h) = \kappa - |h|$. One finds
	\begin{equation}
		I_{\mathrm{SBP}}(x; \sigma) = H\left(-\frac{\kappa + x}{\sigma} \right) - H\left(\frac{\kappa - x}{\sigma} \right)
	\end{equation}
	\item Square Wave (SWP): $\varphi_{\delta}(h) \equiv \mathrm{sign}\left[\sin \left( \frac{\pi}{\delta} h\right)\right]$. The kernel $I_\varphi$ can be efficiently computed by Fourier expansion
	\begin{equation}
		\label{eq:KernelSQWV}
		I_{\mathrm{SWP}}(x ; \sigma) =\frac{1}{2} + \frac{2}{\pi}\sum_{n=0}^{\infty}\frac{\mathrm{e}^{- \frac{\pi^2 \sigma^2}{2\delta^2}\,(2n+1)^2}}{2n+1}\sin{\left(\frac{\pi (2n+1)}{\delta} x\right)}.
	\end{equation}
	Note that for $\lim_{\delta \to \infty}\varphi_\delta(h) = h$, so that the Kernel tends in this limit to the one of the ABP. 
\end{itemize}

\section{On the existence of collisions}
\label{app:alpha_c}
Recall the definitions of the entropy of collisions with internal overlap $q_1$
\begin{equation}
	\label{eq::quenched_entropy_collisions}
	\Phi(q_1) = \lim_{N\to \infty} \frac{1}{N} \mathbb{E}_{\boldsymbol{A}}\log Z (q_1; \boldsymbol{A})
\end{equation}
where
\begin{equation}
	\label{eq::partition_function_existence_collisions}
	Z (q_1; \boldsymbol{A}) \equiv \sum_{\boldsymbol{c}\in \{\pm 1\}^{2N}} \mathbb{X}_{\boldsymbol{A}}(\boldsymbol{c}) \, \prod_{k=1}^m \delta\left(q_i(\boldsymbol{c}) - q_1 \right).
\end{equation}
This section deals with the computation of the annealed and RS upper bounds  $\Phi\le \Phi^{\mathrm{RS}}\le \Phi^{\mathrm{a}}$. As we will see, they coincide. This serves as an indication that the RS estimate is exact $\Phi=\Phi^{RS}$, although a proof would require a second moment calculation, which is outside the scope of this work.

\subsection{First moment computation}
The first moment bound on the existence of collisions can be obtained by computing the so called \emph{annealed} free entropy:
\begin{equation}
	\Phi^{\mathrm{a}}(q_1) \equiv  \frac{1}{N} \log \mathbb{E}_{\boldsymbol{A}} Z (q_1; \boldsymbol{A}) \ge \Phi(q_1).
\end{equation}
This quantity can be directly obtained from the expression of $\phi_m^{\mathrm{a}}(p,q_1)$ in Section~\ref{app::m-clones_computation} by setting $m = 1$. In this limit, the dependence on the overlap $p$ between the $m$ collision clones disappears. One therefore gets for large $N$
\begin{subequations}
	\label{eq::phiA}
	\begin{align}
		\Phi^{\mathrm{a}}(q_1) &= \min_{\hat q_1} \left[G_S(q_1, \hat q_1) + \alpha G_E(q_1) \right] \\
		G_S(q_1, \hat q_1) &= - q_1 \hat q_1 + 2\log 2 + \log \cosh(\hat q_1) \\ 
		G_E(q_1) &= \log  \int \! Dz  \, F_\varphi \left(\sqrt{q_1} z, \sqrt{q_1} z; \sqrt{1-q_1}\right).  
	\end{align}
\end{subequations}
The saddle point equation for $\hat q_1$ can be easily solved:
\begin{equation}
	\hat q_1 = \mathrm{atanh} \, q_1 = \frac{1}{2} \log \left(\frac{1+q_1}{1-q_1}\right)
\end{equation}
which substituted in~\eqref{eq::phiA} gives
\begin{equation}
	\Phi^{\mathrm{a}}(q_1) = \log 2 + H_B(q_1) + \alpha \log  \int Dz  \, F_\varphi \left(\sqrt{q_1} z, \sqrt{q_1} z; \sqrt{1-q_1}\right), 
\end{equation}
$H_B(q_1)$ being the Bernoulli entropy with magnetization $q_1$ defined as
\begin{equation}
	H_B(q_1) = -\frac{1-q_1}{2} \log \left(\frac{1-q_1}{2} \right) - \frac{1+q_1}{2} \log \left(\frac{1+q_1}{2}\right) \,.
\end{equation}
Because of the Markov's inequality the probability that the random variable $Z (\boldsymbol{A}) \ge  1$ is bounded by
\begin{equation}
	P\left[Z (q_1; \boldsymbol{A}) \ge 1 \right] \le \mathbb{E}_{\boldsymbol{A}} Z (q_1; \boldsymbol{A}) = e^{N \Phi^\mathrm{a}(q_1)}.
\end{equation}
The right hand side goes to zero for $N \to 0$ when $\Phi^\mathrm{a}(q_1) < 0$, i.e. when $\alpha \ge \alpha_c^\mathrm{A}(q_1)$ where $\alpha_c^\mathrm{a}(q_1)$ is given by
\begin{equation}
	\alpha_c^\mathrm{a}(q_1) = - \frac{\log(2) + H_B(q_1)}{\log  \int Dz  \, F_\varphi \left(\sqrt{q_1} z, \sqrt{q_1} z; \sqrt{1-q_1}\right)}.
\end{equation}
This means that for $\alpha \ge \alpha_c^\mathrm{a}(q_1)$ with high probability there is no collision. Due to Markov's inequality this is only \emph{upper bound} on the existence of collisions. We analyze below the behavior of $\alpha_c^\mathrm{a}(q_1)$ for two functions $\varphi(h)$ of interest:
\begin{itemize}
	\item Asymmetric Binary Perceptron (ABP): $\varphi (h) = h$. One obtains 
	\begin{equation}
		\alpha_c^\mathrm{a}(q_1) = - \frac{\log 2 + H_B(q_1)}{\log \left(\frac{1}{\pi} \arccos(-q_1)\right) }
	\end{equation}
	In the neighborhood of $q_1 \simeq 1$, this behaves as $\alpha_c^\mathrm{a} \simeq \frac{\pi \log 2}{ \sqrt{ 2 dq_1}}$ where $dq_1 = 1-q_1$. 
	\item Square Wave (SWP): If one fixes $\delta>0$ one numerically finds $\alpha_c^\mathrm{a} \propto \frac{1}{ \sqrt{1-q_1}}$ as $q_1 \to 1$. If one performs first the limit $\delta \to 0$ \emph{before} $q_1 \to 1$ the divergence observed at strictly positive $\delta$ near $q_1 \simeq 1$ is removed. Indeed, since $I(x; \sigma) \to 1/2$ one simply finds
	\begin{equation}
		\alpha_c^\mathrm{a}(q_1) = 1 + \frac{H_B(q_1)}{\log 2} 
	\end{equation}
	which is a monotonically decreasing function of $q_1$. In particular, for $q_1 \to 1$ one finds $\alpha_c^{\mathrm{a}} \to 1$. 
\end{itemize}

\subsection{Upper bound to the existence of Collisions in the Sub-extensive regime}
\label{sec:AnnealedSubExt}
Consider an activation function $\varphi(\bullet)$ that is always $\pm 1$ except for a set of discontinuity points $\{\xi_0,\xi_1,\dots\}$. We assume that the distances between the discontinuity points remain greater than zero in the thermodynamic limit. 

For large $N$, the number of collisions with inputs differing for a single bit flip is given by:
\begin{equation}
	\begin{split}
		\mathbb{E}_{\boldsymbol{A}}Z \left(q_1; \boldsymbol{A}\right)|_{q_1=1-\frac{1}{N}} \approx N\,2^{N} J_{\varphi}^P\approx N\,2^N \left(1-\frac{a_{\varphi}}{\sqrt{N}}\right)^P \,,
	\end{split}
\end{equation}
where
\begin{equation}
	\label{eq:defJphi}
	\begin{split}
		J_{\varphi}=&\int Dz_0Dz_1\Theta\Big(\varphi\left(z_0+z_1/\sqrt{N}\right)\varphi\left(z_0-z_1/\sqrt{N}\right)\Big)\approx 2\int_0^{\infty}Dz_1\,\sum_{n}\int_{\xi_n+z_1/\sqrt{N}}^{\xi_{n+1}-z_1/\sqrt{N}}Dz_0\approx\\&\approx 1-\frac{1}{\sqrt{N}}\frac{2}{\pi}\sum_{n}e^{-\xi_n^2/2},
	\end{split}
\end{equation}
and
\begin{equation}
	\label{eq:defaphi}
	a_{\varphi}=\frac{2}{\pi}\sum_{n}e^{-\xi_n^2/2}.
\end{equation}
Therefore, the annealed free entropy is
\begin{equation}
	\Phi^{\mathrm{a}}=\log{2}-a_{\varphi}\,P\,N^{-3/2},
\end{equation}
which leads to the following estimate for the annealed capacity:
\begin{equation}
	\alpha^{\mathrm{a}}_c=\frac{\log{2}}{a_{\varphi}}N^{1/2}.
\end{equation}

\subsection{The Replica Symmetric Ansatz}

The quenched entropy~\eqref{eq::quenched_entropy_collisions} can be computed by using the \emph{replica trick}
\begin{equation}
	\mathbb{E}_{\boldsymbol{A}} \log Z (q_1; \boldsymbol{A}) = \lim_{m\to0} \frac{1}{m} \log \mathbb{E}_{\boldsymbol{A}} Z^m (q_1; \boldsymbol{A}).
\end{equation}
The computation proceeds by considering $m$ to be an integer, averaging over the disorder $\boldsymbol{A}$ and then analytically continuing the result to $m\to 0$. All the calculations can therefore be reconducted to the one presented in Section~\ref{app::m-clones_computation} for the $m$-cloned entropy with two key differences. Firstly, the number of clones is not fixed to an integer but it is sent to 0. Such an analytical continuation is usually performed by doing an `\emph{ansatz}' over the structure of the overlap matrix $q_{st}^{ab}$. In this section we will employ the same ansatz as in Section~\ref{app::Symmetric_Ansatz}, which in the context of the replica method is called \emph{Replica Symmetric ansatz} (RS). 
Secondly, because of the definition in~\eqref{eq::partition_function_existence_collisions}, there is no constraint on the distance between the clones. As a result, within the RS ansatz, the overlap $p$ should not be fixed but instead treated as a variational parameter to be optimized. 
One therefore finally gets to
\begin{subequations}
	\begin{align}
		&\Phi^{\mathrm{RS}}(q_1) = \max_{p, q_0} \min_{\hat q_0, \hat q_1, \hat p} [G_S(q_0, q_1, p, \hat q_0, \hat q_1, \hat p) + \alpha G_E(q_0, q_1, p)] \\
		&G_S = - q_1 \hat q_1 - p \hat p - q_0 \hat q_0  - (\hat p + \hat q_1 - \hat q_0) + \int D u_1 Du_2 \log \left[2 \cosh\left(\sqrt{2(\hat p - \hat q_0)} \,u_1\right) + 2 e^{2(\hat q_1 - \hat q_0)} \cosh\left(\sqrt{2(\hat p + \hat q_0)}\, u_2\right) \right]\\
		&G_E = \int \prod_s Du_s \log  \int Dz  \, F_\varphi\!\left( a(u_1, u_2, z), a(u_1, -u_2, z); \sigma\right),
	\end{align}
	\label{eq:app_m=0}
\end{subequations}
\noindent where $a(u_1, u_2, z)$ and $\sigma$ are the same quantities defined in~\eqref{eq::convenient_quantities}. 

Interestingly, we have found numerically that the optimization in \cref{eq:app_m=0} always leads to $p=q_0=0$ for any value of $q_1$. One then obtains $\Phi^{\mathrm{RS}}(q_1)=\Phi^{\mathrm{a}}(q_1)$, which gives the same predictions for the threshold corresponding to the existence of collisions $\alpha^{\mathrm{RS}}_c(q_1) = \alpha^{\mathrm{a}}_c(q_1)$. This result does not in principle preclude the possibility that the global optimum of the free entropy in the $m\to 0$ limit requires Replica Symmetry Breaking (RSB). 
However, we have ultimately validated the critical capacity predictions by exhaustive search of the collisions with a given value of $q_1$. Despite the small values of $N$, the numerical results are in very good agreement with the theoretical predictions.

\section{Overlap Gap Transition}

Given an instance of the data matrix $\boldsymbol{A}$, define as $\mathcal{S}_m(p,\epsilon; q_1, \boldsymbol{A})$ the set of all the $m$ configurations $\boldsymbol{c}_1,\dots,\boldsymbol{c}_m \in \{\pm 1\}^{2N}$, such that the following properties hold:
\begin{enumerate}
	\item for each $a\in[m]$, $\boldsymbol{c}_a$ satisfies the constraints~\eqref{eq::constraint_collision}, i.e.\ it is a collision;
	\item each collision $\boldsymbol{c}_a$, $a\in[m]$ has an internal overlap $q_1$;
	\item for each $a\neq b\in[m]$
	\begin{equation}
		p < q_e(\boldsymbol{c}_a, \boldsymbol{c}_b) < p +\epsilon\,.
	\end{equation}
\end{enumerate}
We say that the $m-$OGP property holds if it is possible to find $p$ and $\epsilon$ such that:
\begin{equation}
	\label{eq:CondOGP}
	\text{Pr}_{\boldsymbol{A}}\left[\mathcal{S}_m(p,\epsilon; q_1, \boldsymbol{A})\neq \emptyset\right]\overset{N\uparrow\infty}{\longrightarrow} 0\,.
\end{equation}
We call by $\alpha_{\mathrm{OGP}}(m ; q_1)$ the constraint density above which the $m$-OGP begins to hold for collisions constrained to have an internal overlap $q_1$.

The general procedure to be adopted to obtain the $m$-OGP transition from our analytical framework is the following. Fix the value of $\alpha$, $p$ and $q_1$. Solve the corresponding maximization procedure of equation~\eqref{eq::max_procedure}. Then impose the following additional constraints on the values of $p$ and $\alpha$
\begin{subequations}
	\begin{align}
		&p^\star_m(\alpha; q_1) = \mathrm{argmin}_{p} \phi_m(p, q_1) \\
		&\phi_m(p^\star_m(\alpha; q_1), q_1) = 0.
	\end{align}
\end{subequations}
Namely, the value of $p_m^\star$ at the $m-$OGP transition represents the value of external overlap at which $m$ collisions should be disposed to have zero entropy. For $\alpha > \alpha_{\mathrm{OGP}}(m; q_1)$, therefore, since the entropy is a decreasing function of $\alpha$, there exists a region of external overlap between collisions $p$ around $p^\star_m(\alpha; q_1)$ for which the entropy is negative. In other words, it is possible to find $\epsilon>0$ s.t.\ condition \eqref{eq:CondOGP} is satisfied, signaling the presence of a \emph{gap} in the overlaps. 

We stress here that our procedure is only guaranteed to give an upper bound to the true value of the OGP transition, as we are using a first moment to estimate the entropy of the $m-$cloned system. 

We also underline that the maximization in equation~\eqref{eq::max_procedure} should be carried over a generic structure of the admissible correlation matrices $Q$. This procedure is difficult to deal analytically for generic non-linearities $\varphi$ and number of clones $m$. The next subsections are therefore organized as follows. In section~\ref{app::2OGP_RHF} we compute the 2-OGP transition in the SHL by computing the global maximum of the saddle point equations. We will see that this maximum is actually attained by a correlation matrix which satisfies the symmetric ansatz introduced in~\ref{app::Symmetric_Ansatz}. Finally in section~\ref{app::Symmetric_Ansatz_RHF} we restrict the maximization procedure on matrices $Q$ satisfying the symmetric ansatz, and find the upper bound to the OGP threshold for generic $m$ for the SHL. The same procedure has been applied to compute the $m$-OGP threshold at finite $\delta$. 

\subsection{2-OGP in the SHL}~\label{app::2OGP_RHF}

We start here to examine the emergence of 2-OGP in the SHL i.e.\ for $\delta \to 0$. We will compute the 2-OGP threshold for an arbitrary structure of the covariance matrix $Q$. 

We shall decompose $\mathcal{F}(p, q_1)$ according to the number of 0 eigenvalues: we write
\begin{equation}
	\mathcal{F}(p, q_1) = \bigcup_{k=0}^{2m-1} \mathcal{F}_{k}(p, q_1)
\end{equation}
where $\mathcal{F}_{k}(p, q_1)$ the set of admissible matrices with exactly $k$ eigenvalues equal to zero, the other $2m - k$ eigenvalues being larger than zero. We do not include $\mathcal{F}_{2m}(p, q_1)$ as this is obviously an empty set. We therefore write equation~\eqref{eq::max_procedure} as
\begin{align}
	\phi_m^\star(p, q_1) &= \max_{k\in\{0\,, \dots\,, 2m-1\}} G_m^{(k)}(p, q_1) \\
	G_m^{(k)}(p, q_1) &= \max_{Q \in \mathcal{F}_k(p, q_1)} \phi_m(Q).
\end{align} 
In the next subsections, we study the case $m=2$ in detail, analyzing the form of $Q_k^\star$, i.e.\ the maximum on each of the sets $\mathcal{F}_k$. In the $m=2$ case we parametrize the covariance matrix $Q$ by
\begin{equation}
	\label{eq::Q_parametrization}
	Q = 
	\begin{pmatrix}
		1 & q_1 & r_{xx} & r_{xy} \\
		q_1 & 1 & r_{yx} & r_{yy} \\
		r_{xx} & r_{yx} & 1 & q_1 \\
		r_{xy} & r_{yy} & q_1 & 1
	\end{pmatrix}
\end{equation}
with the fixed distance over collisions that imposes $r_{xx} + r_{yy} = 2p$. 

\subsubsection{The maximum on the $\mathcal{F}_0$ set}
When all the eigenvalues of $Q$ are positive, in the strong hashing limit we have 
\begin{equation}
	G_E(Q^\star_{0}) \equiv G_E^{k=0} = - \frac{1}{2}\log 2
\end{equation}
for all admissible $Q$. One has to find $Q$ by maximizing the $G_S(Q)$ term only. The maximum is found at
\begin{equation}
	Q_{0}^\star(p, q_1) = 
	\begin{pmatrix}
		1 & q_1 & p & q_0 \\
		q_1 & 1 & q_0 & p \\
		p & q_0 & 1 & q_1 \\
		q_0 & p & q_1 & 1
	\end{pmatrix}
\end{equation}
with a conjugate matrix 
\begin{equation}
	\hat Q_{0}^\star(p, q_1) = 
	\begin{pmatrix}
		0 & \hat q_1 & \hat p & 0 \\
		\hat q_1 & 0 & 0 & \hat p \\
		\hat p & 0 & 0 & \hat q_1 \\
		0 & \hat p & \hat q_1 & 0
	\end{pmatrix}
\end{equation}
where $\hat p$, $\hat q_1$ and $q_0$ are functions of $p$ and $q_1$ that can be obtained by solving the following coupled equations
\begin{subequations}
	\label{eq::F0_hat_parameters}
	\begin{align}
		p &= \frac{\cosh (2 \hat q_1) \sinh (2 \hat p)}{1 + \cosh (2 \hat p) \cosh (2 \hat q_1)}\\
		q_1 &= \frac{\cosh (2 \hat p) \sinh (2 \hat q_1)}{1 + \cosh (2 \hat p) \cosh (2 \hat q_1)} \\
		q_0 &= \frac{\sinh (2 \hat q_1) \sinh (2 \hat p)}{1 + \cosh (2 \hat p) \cosh (2 \hat q_1)}
	\end{align}
\end{subequations}
Note that $Q_0^\star$ is a matrix that shares a structure similar to that obtained under a Symmetric ansatz, see section~\ref{app::Symmetric_Ansatz}.

Let's show that this is a maximum. We first notice that 
\begin{equation}
	\frac{\partial G_S}{\partial Q_{\alpha \beta}} = \hat Q_{\alpha \beta} \,,
\end{equation}
which implies
\begin{equation}
	\frac{\partial G_S}{\partial Q_{14}} = \frac{\partial G_S}{\partial Q_{23}} = 0 \,.
\end{equation}
Then we also have 
\begin{equation}
	\frac{\partial G_S}{\partial Q_{13}} = \frac{\partial G_S}{\partial Q_{24}} = \hat p
\end{equation}
Writing $Q_{13} = r$ and $Q_{24} = 2p - r$ we have
\begin{equation}
	\left. \frac{\partial G_S}{\partial r} \right|_{r = p} = \frac{\partial G_S}{\partial Q_{13}} - \frac{\partial G_S}{\partial Q_{24}} = 0
\end{equation}
Therefore $Q_0^\star$ is a maximum. As $G_S$ is concave, this is the maximum. \eqref{eq::F0_hat_parameters} can be found by taking the expression found for the entropic term $G_S$ in the Symmetric ansatz, see section~\ref{app::Symmetric_Ansatz}, and imposing vanishing derivatives with respect to $\hat p$, $\hat q_1$ and $q_0$, with $m=2$ and $\hat q_0 = 0$. 

\subsubsection{The maximum on the $\mathcal{F}_1$ set}

If one eigenvalue of $Q$ vanishes, the integration measure $\langle \bullet \rangle$ in ~\eqref{eq:Ge_form1} is constrained on the subspace orthogonal to the zero mode. This means that it exists one linear combination of the variable of integration imposing such a constraint, i.e.
\begin{equation}
	a u_1 +b v_1 +c u_2 +d v_2 = 0 \,.
\end{equation}
A dangerous case could be $c = d = 0$ (or equivalently $a=b=0$), then $v_1 = t u_1$: for some values of $t$ the hashing in $u_1$ and $v_1$ is in phase and the integral $\int du_1 \Theta( \varphi(u_1) \varphi (t u_1))$ turns out to be different from $1/2$, even in the strong hashing limit (for example, take the easy case $t=1$). However we note that, if $v_1 = t u_1$ the fact that $\langle (v_1)^2\rangle = \langle(u_1)^2\rangle = 1$ implies that $t = \pm 1$. But then the internal overlap $q_1 = \langle u_1 v_1\rangle = \pm 1$, which is violated by our constraint on $q_1$.

The other potentially dangerous case is $v_2 = a u_1 + b v_1 + c u_2$. The constraint in~\eqref{eq::Q_zero_eigenvalue_contraint} then imposes that there exist three spins $\tau_1, \tau_2, \tau_3 = \pm 1$ such that $c = \tau_1 + a \tau_2 + b \tau_3$. One can search the most dangerous case by computing the small $\delta$ limit of
\begin{equation}
	I(a, b) = \max_{\tau_1, \tau_2, \tau_3} \int Du_1 Du_2 Dv_1 Dv_2 \, \Theta\left(\varphi(u_1) \varphi(v_1)\right) \Theta\left(\varphi(u_2) \varphi(v_2)\right) \delta\left(v_2 - a u_1 - b v_1 - (\tau_1 + a \tau_2 + b \tau_3) u_2\right).
\end{equation}
Note that here we use independent Gaussian for $u_1$, $u_2$, $v_1$. In principle we should use the covariance matrix of these three
variables, but as long as this covariance matrix is not singular, in the strong hashing limit $\delta \to 0$ the result is the same.

\begin{figure}[ht]
	\centering
	\includegraphics[width=0.49\linewidth]{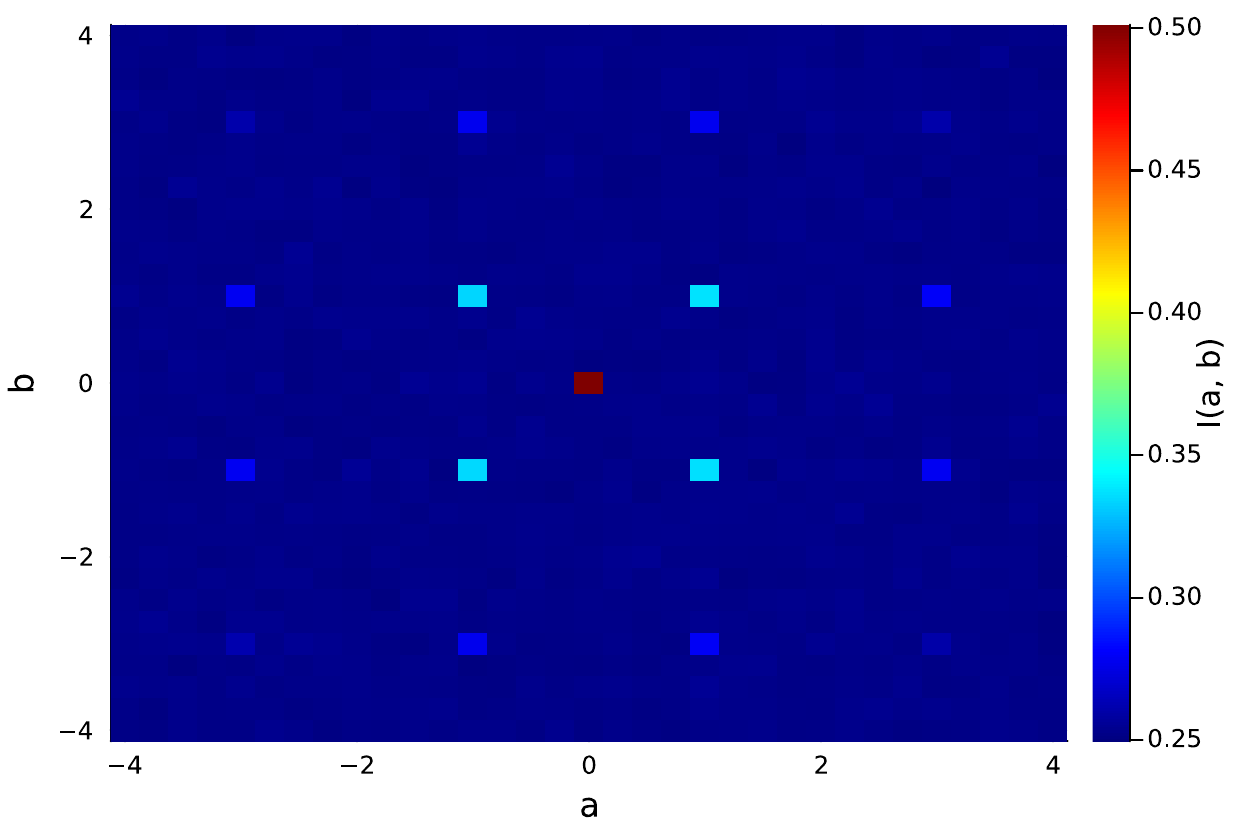}
	\includegraphics[width=0.49\linewidth]{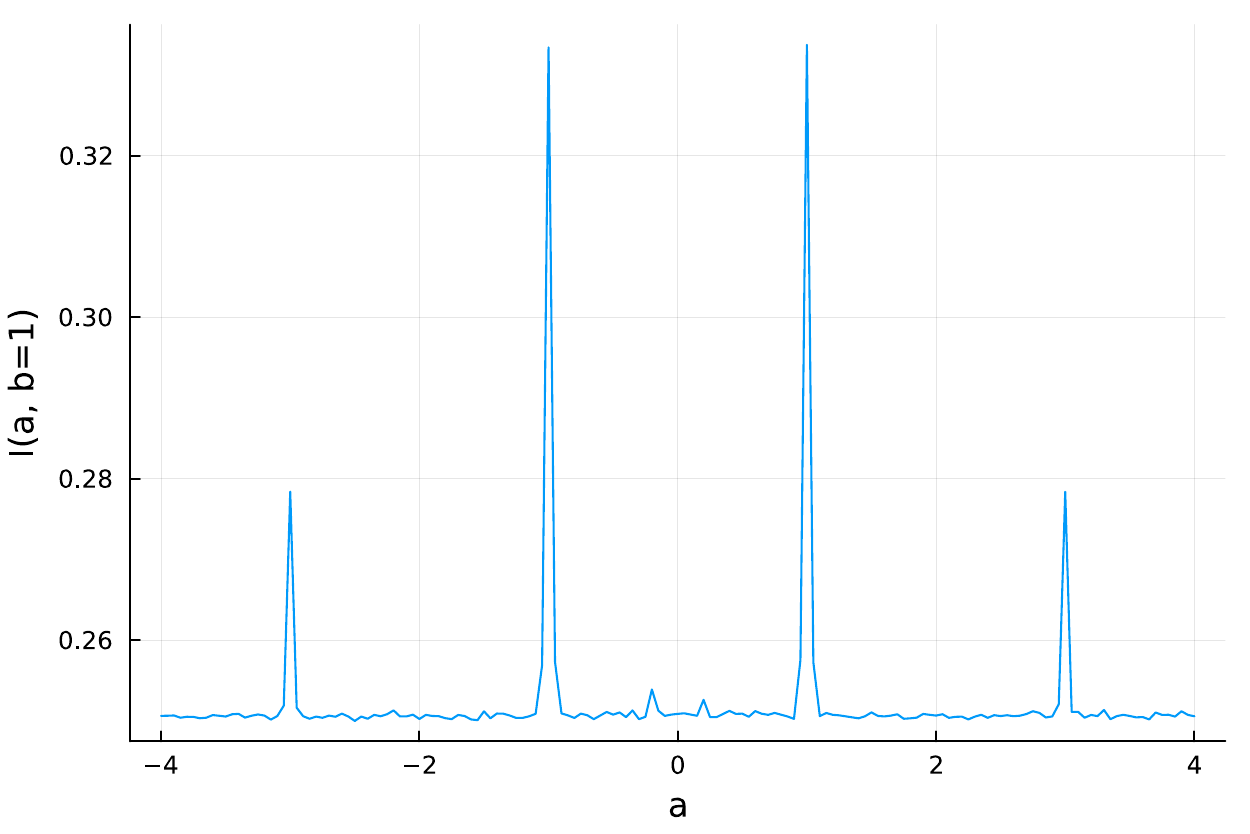}
	\caption{Left: heatmap plot of $I(a, b)$. The function is equal to 1/4 almost everywhere except at some discrete value of $a$ and $b$. Right: plot of $I(a, b)$ as a function of $a$ with $b=1$. $I(a, b)$ is evaluated by sampling $10^6$ random Gaussian variables $u_1$, $u_2$ and $u_3$ for each of the possible combinations of $\tau_1$, $\tau_2$ and $\tau_3$ and then computing the maximum over those variables. Here we have also considered $\delta = 0.1$. Note that $I(a, b) = 1/2$ for $a=b=0$ and $1/3$ for $a= \pm 1$ and $b=\pm 1$. }
	\label{fig:I(a,b)}
\end{figure}
A numerical study of this integral, see Figure~\ref{fig:I(a,b)}, allows us to conjecture that the maximum value is obtained
when $a = \pm 1$, $b = \pm 1$ (the global maximum located at $a=b=0$ is again not admissible as examined before). By symmetry we can study one of these four cases, say $a = -1$, $b = 1$, $c = 1$. The singular eigenvalue enforces the constraint $v_2 = u_2 + v_1 - u_1$ or 
\begin{equation}
	\label{eq::contraint_F1}
	v_2 - u_2 = v_1 - u_1.
\end{equation}
We can compute the value of the energetic term corresponding to $Q_1^\star$. In this case calling $g=u_1 - v_1$ one finds
\begin{equation}
	I(-1, 1) = \int Dg \, \mathcal{K}^2(g)
\end{equation}
with
\begin{equation}
	\mathcal{K}(g) \equiv \int Du \, \Theta(\varphi(u) \varphi(u+g)).
\end{equation}
The function $\mathcal{K}(g)$ is periodic of period $2\delta$, and its value in the interval $[-\delta, \delta]$ is $K(g) = 1 - \frac{|g|}{\delta}$. Therefore $I(-1, 1) = 1/3$, and
\begin{equation}
	G_E^{k=1} = \frac{1}{2} \log I (-1, 1) = - \frac{1}{2}\log 3.
\end{equation}
We need to check whether there exists an admissible covariance matrix with this singular eigenvector, and what is the entropy associated with this matrix. We use the parametrization in~\eqref{eq::Q_parametrization} for the covariance matrix. The constraint
$v_2 = u_2 + v_1 - u_1$ imposes that the elements in the last column $r_{xy} = \langle u_1 v_2\rangle$, $r_{yy} = \langle v_1 v_2 \rangle$, $q_1 = \langle u_2 v_2\rangle$, $\langle (v_2)^2 \rangle = 1$ of the matrix can be expressed in terms of $r_{xx}$, $r_{yx}$. This gives:
\begin{subequations}
	\begin{align}
		r_{xy} &= r_{xx} + q_1 - 1 \\
		r_{yy} &= r_{yx} + 1 - q_1 \\
		q_1 &= 1+ r_{yx} - r_{xx} \\
		1 &= q_1 + r_{yy} - r_{xy}
	\end{align}
\end{subequations}
which give
\begin{subequations}
	\begin{align}
		r_{xx} &= r_{yy} = p \\
		r_{xy} &= r_{yx} = p + q_1 - 1.
		\label{eq::non_trivial_q0_condition}
	\end{align}
\end{subequations}
Altogether, we have the following family of dangerous admissible covariance matrices parametrized by $q_1$ and $p$ which have one eigenvalue equal to 0, implementing the constraint in~\eqref{eq::contraint_F1}
\begin{equation}
	Q_{1}^\star(p, q_1) = 
	\begin{pmatrix}
		1 & q_1 & p & p+q_1-1 \\
		q_1 & 1 & p+q_1-1 & p \\
		p & p+q_1-1 & 1 & q_1 \\
		p+q_1-1 & p & q_1 & 1
	\end{pmatrix}
\end{equation}
which again has the same structure of the ones of the symmetric ansatz.

\subsubsection{The maximum on the $\mathcal{F}_2$ set}

When two eigenvalues are zero, we integrate over a plane in the space $u_1$, $v_1$, $u_2$, $v_2$. We first study the case when this plane is in a generic direction, so that
\begin{subequations}
	\label{eq::F2_contraints_planes_generic_direction}
	\begin{align}
		a u_1 + b v_1 - u_2 &= 0 \\
		c u_1 + d v_1 - v_2 &= 0.
	\end{align}
\end{subequations}
Using $\langle (u_1)^2\rangle = \langle (v_1)^2\rangle = 1$ and $\langle (u_1)\rangle = q_1$ we get
\begin{subequations}
	\begin{align}
		r_{xx} &= a + b q_1 \\ 
		r_{yy} &= c q_1 + d \\
		r_{xy} &= c + d q_1 \\
		r_{yx} &= a q_1 + b \\
		\label{eq::F2_useful_equation3}
		q_1 &= ac + bd + (bc + ad) q_1 \\
		\label{eq::F2_useful_equation}
		1 &= a^2 + b^2 + 2 ab q_1 \\
		\label{eq::F2_useful_equation2}
		1 &= c^2 + d^2 + 2cd q_1.
	\end{align}
\end{subequations}
We also need to implement the constraints in~\eqref{eq::Q_zero_eigenvalue_contraint} for each of the eigenvectors with zero eigenvalue. In the present case, there are two such eigenvectors characterized by~\eqref{eq::F2_contraints_planes_generic_direction}. Let us start with the first one, i.e.\ $(a, b, -1, 0)$; there should exist $\tau_1$, $\tau_2 \in \pm 1$ such that $a = \tau_1 + b \tau_2$. Using~\eqref{eq::F2_useful_equation} one finds that either $b=0$ and $a=\pm 1$ which implies $r_{xx} = \pm 1$, $r_{yx} = \pm q_1$, or $a=0$, $b=\pm 1$ which gives $r_{xx} = \pm q_1$, $r_{yx} = \pm 1$. 

A similar analysis can be carried out for the second eigenvector $(c, d, 0, -1)$. Using~\eqref{eq::F2_useful_equation2} one finds that either $c=0$, $d=\pm 1$ which gives $r_{yy}= \pm 1$, $r_{xy}=\pm q_1$ or $d=0$, $c=\pm 1$ which gives $r_{yy} = \pm q_1$ and $r_{xy}=\pm 1$. Putting those results together and using~\eqref{eq::F2_useful_equation3} one finds either that $q_1=\pm 1$, or that $p$ needs to assume special values $p=1$ which are forbidden.

Planes in a non-generic direction are parallel to some of the axes, but then they need to have at least one of the components which vanishes, say $u_1 = 0$. But this is excluded by $\langle (u_1)^2 \rangle = 1$. Therefore $\mathcal{F}_2 = \emptyset$.

\subsubsection{The maximum on the $\mathcal{F}_3$ set}
When three eigenvalues are zero, we integrate over a line. We can parametrize it as:
\begin{equation}
	u_1 = t\,, \quad v_1 = a t\,, \quad u_2 = bt\,, \quad v_2 = ct    
\end{equation}
$\langle (u_1)^2\rangle = \langle (v_1)^2\rangle = 1$ implies that $a=\pm 1$, but then we get $q_1 = a = \pm 1$ which is excluded. Therefore $\mathcal{F}_3 = \emptyset$.

\subsubsection{The global maximum}

In order to study the global maximum, we have to compare the entropy attained by using the two matrices $Q^\star_{0}$ and $Q_{1}^\star$ respectively obtained imposing zero and one vanishing eigevalues. 
Notice that $G_E^{k=0} \ge G_E^{k=1}$ but since $r_{xy} = r_{yx} \ne 0$ for $Q^\star_{1}$, the entropic terms $G_S^{k=0} \le G_S^{k=1}$. However, in order to find the value of $\alpha_{\mathrm{OGP}}(q_1)$ one has to find the value of $p$ where the entropy is minimal. This is attained in both the $k=0,1$ cases when $p\to 1$. This should be expected, as the entropic term is always a monotonically decreasing function of the overlap between collisions $p$; indeed if one constraints more the distance (namely $p$ is larger) one has less entropy. When there is no dependence on the energetic term on $p$, therefore, the minimal entropy is found for $p \to 1$. In this limit one finds
\begin{equation}
	Q_{0}^\star = Q_{1}^\star = 
	\begin{pmatrix}
		1 & q_1 & p & q_1 \\
		q_1 & 1 & q_1 & p \\
		p & q_1 & 1 & q_1 \\
		q_1 & p & q_1 & 1
	\end{pmatrix}
\end{equation}
Indeed for $k=0$ equations~\eqref{eq::F0_hat_parameters} in the limit $p\to1$ give $\hat p \to \infty$, $\hat q_1 = \frac{1}{2} \mathrm{arctanh}(q_1)$ and $q_0 = \tanh(2 \hat q_1) \to q_1$. Therefore the entropic terms become equal in this limit to 
\begin{equation}
	\lim_{p\to 1} G_S^{k=0}(p, q_1) = \lim_{p\to 1} G_S^{k=1}(p, q_1) = \frac{1}{2}\log(2) + \frac{1}{2} H_B(q_1) 
\end{equation}
Therefore the global maximum in the $p\to 1$ is given by covariance matrices $Q \in \mathcal{F}_1$. The $2-$OGP transition is finally given by 
\begin{equation}
	\lim_{p \to 1} \phi_{m=2}^\star(p, q_1) = 0 \implies \alpha_{\mathrm{OGP}}(m=2; q_1) = 
	\frac{ \log(2) + H_B(q_1) }{\log 3}
\end{equation}

\subsection{Analysis of the SHL in the symmetric ansatz for generic $m$} \label{app::Symmetric_Ansatz_RHF}

\subsubsection{Hubbard-Stratonovich}
Here we compute the energetic term in the symmetric ansatz for generic $m$, in the limit $\delta \to 0$. We will use the Hubbard-Stratonovich representation given in~\eqref{eq::Ge_useful} that we rewrite for convenience:
\begin{equation}
	G_E =\frac{1}{m}\log \int \prod_s Du_s Dx \left[ \int Dz \prod_{s} D h_s \, \Theta\left( \prod_s \varphi(\sqrt{1-q_1 - p + q_0} h_s + \sqrt{p-q_0} u_s + \sqrt{q_0} x + \sqrt{q_1 - q_0}z)\right) \right]^m.
	\label{eq::Ge_Hubbard_again}
\end{equation}
Note that the term $1-q_1 -p + q_0$ appears explicitly as one of the terms under the square roots. When this quantity vanishes, it gives exactly the condition on $q_0$ that was explicitly derived for the case $m=2$, see~\eqref{eq::non_trivial_q0_condition}, and that was related to the presence of a singular eigenvalue in the covariance matrix $Q$, as given in~\eqref{eq::contraint_F1}. The case of $\sigma^2 = 1-q_1 -p + q_0 \sim 0$ must therefore be handled with care.

If $1-q_1 - p + q_0>0$, we can actually use the $G_E$ given in~\eqref{eq:Ge_general} with the kernel $I(m,\sigma)$ given by~\eqref{eq:KernelSQWV}. As $\sigma^2 = 1-q_1 -p + q_0>0$ all the terms in the series vanish exponentially fast and give $\lim_{\delta \to 0} I_{\mathrm{SWV}}(m,\sigma) = 1/2$. This leads to a constant $G_E$
\begin{equation}
	G_E=-\log{2}, \quad 1-q_1 - p + q_0>0.   
\end{equation}
When $1-q_1 - p + q_0 = 0$, the determinant of $Q$ vanishes, the Gaussian measure in~\eqref{eq:Ge_form1} becomes singular, and the integration is effectively reduced to an integration over a lower-dimensional space. Plugging into~\eqref{eq::Ge_Hubbard_again} the expression $q_0=q_1 + p -1$, one obtains
\begin{equation}
	G_E =\frac{1}{m}\log \int \prod_s Du_s Dx \left[ \int Dz \, \Theta\left( \prod_s \varphi( m_s+ \sqrt{1-p}z)\right) \right]^m \quad \text{where } m_s:= \sqrt{1-q_1}u_s+\sqrt{p+q_1-1}\,x.
\end{equation}
The Kernel $\mathcal{K}(m_1, m_2):= \int Dz \, \Theta\left( \prod_s \varphi( m_s + \sqrt{1-p}\,z)\right)$ is translationally invariant due to the periodicity of the SWP activation and therefore depends only on $g:=m_2-m_1$. Moreover, it is periodic in $g$ with period $2\delta$. Defining $\mathcal{K}(g):= \int Dz \, \Theta\left(\varphi(\sqrt{1-p}z)\varphi(g+ \sqrt{1-p}z)\right)$, one has 
\begin{equation}
	G_E =\frac{1}{m}\log \int D g \, \mathcal{K}^m\left(\sqrt{2(1-q_1)}g \right).
	\label{eq:short_ge}
\end{equation}
Within the Kernel, the $\Theta$ function is also periodic in $z$, with period $\delta$. To compute the Kernel in the small $\delta$ limit, it is convenient to expand it in Fourier series
\begin{equation}
	\Theta\left(\varphi(t)\varphi(g+ t)\right)=\frac{1}{2} a_0(g)+\sum_{i=1}^\infty a_i(g) \cos\left(\frac{2\pi n}{\delta}t\right)+\sum_{i=1}^\infty b_i(g) \sin\left(\frac{2 \pi n}{\delta}t\right),
\end{equation}
where
\begin{subequations}
	\begin{align}
		a_0(g) &=\frac{2}{\delta}\int_{0}^{\delta} \Theta\left(\varphi(t)\varphi(g+ t)\right) dt\,, \\
		a_i(g) &=\frac{2}{\delta}\int_{0}^{\delta} \Theta\left(\varphi(t)\varphi(g+ t)\right) \cos\left(\frac{2 \pi n}{\delta} t\right) dt \,, \\
		b_i(g) &=\frac{2}{\delta}\int_{0}^{\delta} \Theta\left(\varphi(t)\varphi(g+ t)\right) \sin\left(\frac{2\pi n}{\delta} t\right) dt.
	\end{align}
\end{subequations}
Upon integration with the Gaussian, only the zero mode gives a finite contribution in the $\delta\to0$ limit, so that
\begin{equation}
	\mathcal{K}(g)= \int Dz \, \Theta\left[\varphi\left(\sqrt{1-p} z\right)\varphi\left(g + \sqrt{1-p} z\right)\right] = \frac{1}{2} a_0(g) = 1-\frac{|g|}{\delta}\,,  \qquad g \in [-\delta, \delta] \,.
\end{equation}
This profile repeats periodically with a period $2\delta$.  
The Gaussian integral in \cref{eq:short_ge} can be similarly computed by decomposing $\mathcal{K}(g)$ in Fourier modes, and keeping only the zero mode. This leads to a different constant value of $G_E$ in the $\delta \to 0$ limit, for points on the boundary $1-q_1 - p + q_0=0$:
\begin{equation}
	\label{eq::Ge_final_result_symmetric}
	G_E=-\frac{1}{m}\log (1+m), \quad 1-q_1 - p + q_0=0. 
\end{equation}
Since both the energetic term are going to a constant, the maximum of the entropy is attained for values of the overlaps on the boundary $1-q_1 - p + q_0 = 0$. 
The optimum of the entropic contribution $G_I+G_S$ is found at $p=1$, implying that $\hat p \to \infty$, $q_0 \to q_1$ and $\hat q_0 \to \hat q_1$, giving
\begin{equation}
	G_S = - m q_1 \hat q_1 + \frac{2 \log 2}{m}+ \frac{1}{m} \log \cosh(m^2 \hat q_1). 
\end{equation}
One can then find the solution of $\hat q_1$ as a function of $q_1$
\begin{equation}
	\hat q_1 = \frac{1}{m^2} \,  \mathrm{atanh}(q_1) =  \frac{1}{2m^2} \log \left(\frac{1+q_1}{1-q_1}\right).
\end{equation}
Putting all together, the result of the optimization in the $\delta\to0$ limit is
\begin{equation}
	\phi_m(q_1) = \frac{\log2 + H_B(q_1)}{m}-\alpha\frac{\log(1+m)}{m}
\end{equation}
which gives 
\begin{equation}
	\alpha_{\mathrm{OGP}}^m(q_1) = \frac{\log 2 + H_B(q_1)}{\log(1+m)}
\end{equation}
which is monotonically decreasing in $m$ and vanishes for $m \to \infty$ for every $q_1$. 

\subsubsection{Fourier representation}

In this section, we show that in the SHL the same result for the energetic term~\eqref{eq::Ge_final_result_symmetric} that was derived from the Hubbard-Stratonovich representation, can be obtained starting from its Fourier representation~\eqref{eq:Ge_fourier}, which
covers all the domain where the covariance matrix $Q$ is positive definite. In the symmetric ansatz, the Fourier representation of the energetic term~\eqref{eq:Ge_fourier} becomes
\begin{equation}
	\label{eq:Ge_fourier_symmetric}
	G_E = \frac{1}{m} \ln \left[ \frac{1}{2^m} + \frac{1}{2^m}   \sum_{r=1}^m \binom{m}{r} \left(-\frac{4}{\pi^2}\right)^r \sum_{\boldsymbol{n} \in \mathbb{Z}_{\text{odd}}^{2r}} \frac{1}{\prod_{i=1}^{2r} n_i} \, e^{- \frac{\pi^2}{2\delta^2} \boldsymbol{n}^T Q_r \boldsymbol{n}}  \right],
\end{equation}
where $Q_r$ is the covariance matrix of $r$ clones in the symmetric ansatz and we have denoted by $\mathbb{Z}_{\text{odd}}^{2r}$ the set $\mathbb{Z}^{2r}$ where each entry is an odd number. In the strong hashing limit $\delta \to 0$, for each $r$ the sum over $\boldsymbol{n}$ is dominated by the eigenvectors of $Q_r$. When $r=1$, $Q_{r=1}$ has eigenvalues $1-q_1$ and $1+q_1$. Since we are assuming $1<q_1<1$, the $r=1$ will always vanish in the strong hashing limit.

For $r>1$, $Q_r$ has eigenvalues
\begin{subequations}
	\begin{align}
		\lambda_1 &= 1 - p + q_0 - q_1\,, \qquad\qquad \qquad\qquad\qquad d_1 = r-1 \\
		\lambda_2 &= 1 - p - q_0 + q_1\,, \qquad\qquad\qquad\qquad\qquad d_2 = r-1 \\
		\lambda_3 &= 1 + (r-1) p - (m-1) q_0 - q_1\,, \qquad\; \; \, \quad d_3 = 1 \\
		\lambda_4 &= 1 + (r-1) p + (m-1)q_0 + q_1\,, \qquad \; \; \,  \quad d_4 = 1.
	\end{align}
\end{subequations}
We call the corresponding eigenvectors respectively as $\boldsymbol{\eta}^{1, s}$, $\boldsymbol{\eta}^{2, s}$, with $s\in [r-1]$ and $\boldsymbol{\eta}^3$ and $\boldsymbol{\eta}^4$. Their expression is
\begin{subequations}
	\begin{align}
		\eta^{1, s}_i &= \begin{cases}
			(-1)^{i+1} & i = 1, 2, 2s+1, 2s+2\\
			(-1)^{i}   & i = 2s+1, 2s+2\\
			0 & \text{otherwise}
		\end{cases}\\
		\eta^{2, s}_i &= \begin{cases}
			-1 & i = 1, 2 \\
			1 & i = 2s+1, 2s+2\\
			0 & \text{otherwise}
		\end{cases}\\
		\eta^3_i &= (-1)^i\,, \qquad i \in [2r] \\
		\eta^4_i &= 1 \,, \qquad i \in [2r].
	\end{align}
\end{subequations}
Depending on which eigenvalue vanishes, the energetic term goes to a different finite limit as $\delta \to 0$. We therefore analyze the contribution of each eigenvalue and finally select the one that gives the maximum contribution. 

We start by considering the non-degenerate eigenvalues, which are easier to analyze. If we suppose for example that $\lambda_3 = 0$, the index $\boldsymbol{n} \in \mathbb{Z}^{2r}_{\text{odd}}$ can be written as $\boldsymbol{n} = k \boldsymbol{\eta}^3$, where $k \in \mathbb{Z}_{\text{odd}}$. The argument of the series~\eqref{eq:Ge_fourier_symmetric} is therefore $\prod_{i=1}^{2r} n_i = k^{2r} \prod_{i=1}^{2r} (-1)^{i} = k^{2r}$. It is easy to see that the case $\lambda_4 = 0$, gives the same result. 

Therefore if either $\lambda_3$ or $\lambda_4$ vanish, in the SHL one obtains
\begin{equation}
	\lim_{\delta \to 0} \sum_{\boldsymbol{n} \in \mathbb{Z}_{\text{odd}}^{2r} } \frac{1}{\prod_{i=1}^{2r} n_i} \, e^{- \frac{\pi^2}{2\delta^2} \boldsymbol{n}^T Q_r \boldsymbol{n}} = \sum_{k \ne 0} \frac{1}{(2k +1)^{2r}} = 2(1 - 4^{-r}) \zeta(2r),
\end{equation}
where $\zeta$ is the Riemann zeta function. The energetic term reads in this case as 
\begin{equation}
	\label{eq::GE_lambda3_4}
	\lim_{\delta \to 0} G_E(q_1) = \frac{1}{m} \ln \left[ \frac{1}{2^m} + \frac{1}{2^{m-1}}   \sum_{r=2}^m \binom{m}{r} \frac{(-1)^r (4^r - 1)}{\pi^{2r}} \zeta(2r) \right].
\end{equation}
Let us now continue analyzing just the case $\lambda_1 = 0$. The only difference with respect to the non-degenerate eigenvalues is that now the index $\boldsymbol{n}$ can be written as a linear combination of the $r-1$ eigenvectors:
\begin{equation}
	\boldsymbol{n} = \sum_{s=1}^{r-1} k_s \boldsymbol{\eta}^{3, s}
\end{equation}
$k_s$, $s \in [r-1]$ should be chosen in such a way that \emph{every entry} in $\boldsymbol{n}$ is an odd number. Due to the structure of the $r-1$ eigenvectors $\boldsymbol{\eta}^{3,s}$ it is easy to see that we have to use at least once each eigenvector, namely $k_s \ne 0$, $\forall s \in [r-1]$. 

If $r$ is odd, however it is impossible to satisfy this constraint and at the same time to maintain all the components of $\boldsymbol{n}$ even. We therefore conclude that if $r$ is odd the contribution to the sum always vanishes. 

\begin{figure}[ht]
	\centering
	\includegraphics[width=0.4\linewidth]{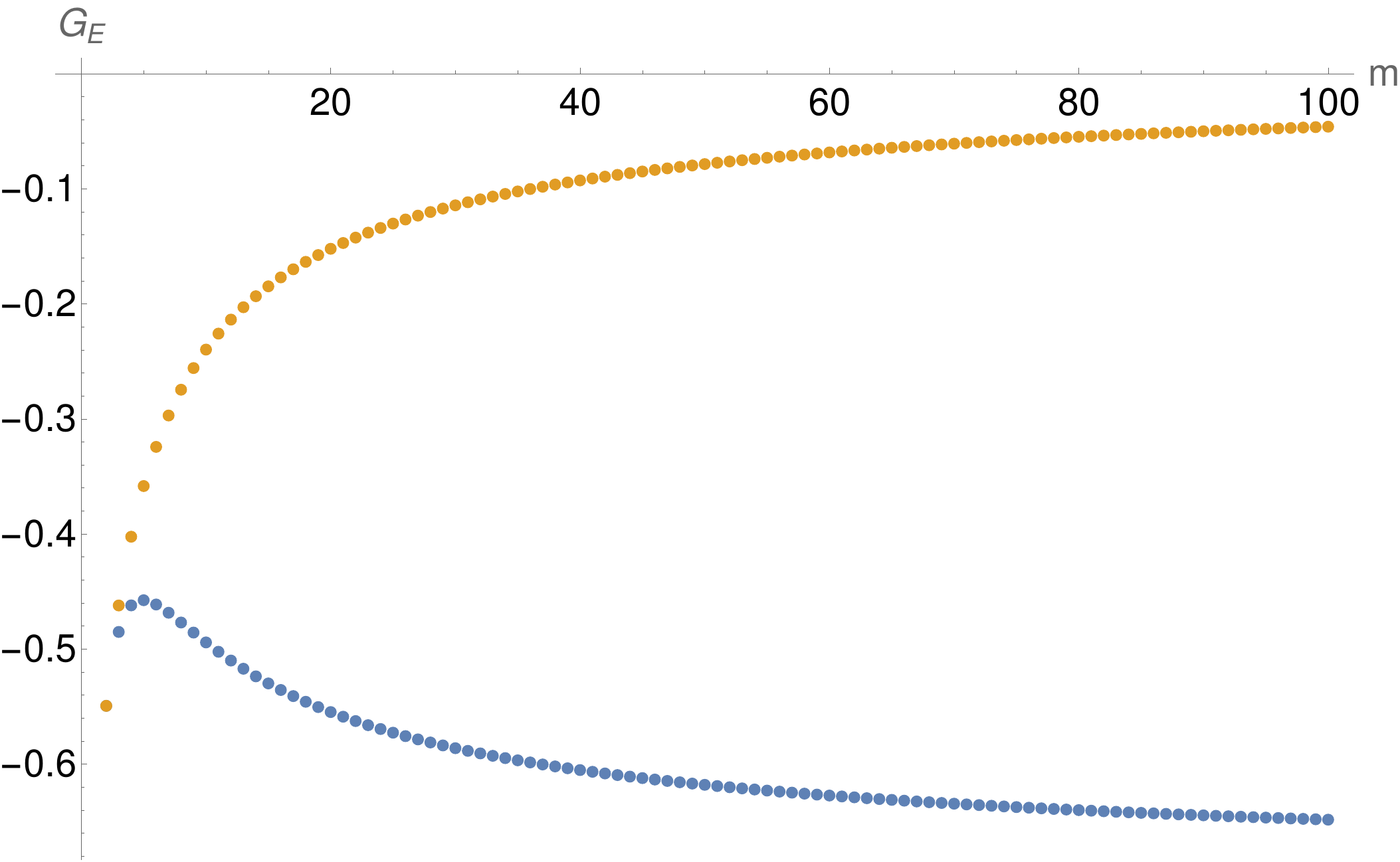}
	\caption{Energetic term corresponding to the vanishing of $\lambda_1$ or $\lambda_2$ (orange) and to the vanishing of $\lambda_3$ or $\lambda_4$  (blue) as a function of $m$. Note that for $m=2$ is the only case where all the contributions coincide.}
	\label{fig:Ge_Fourier}
\end{figure}

If $r$ is even the constraint that $\boldsymbol{n} \in \mathbb{Z}^{2r}_{\text{odd}}$ simply imposes that $k_s$ is odd for all $s \in [r-1]$. If $r$ is even, then we get
\begin{equation}
	\lim_{\delta \to 0} \sum_{\boldsymbol{n} \in \mathbb{Z}_{\text{odd}}^{2r} } \frac{1}{\prod_{i=1}^{2r} n_i} \, e^{- \frac{\pi^2}{2\delta^2} \boldsymbol{n}^T Q_r \boldsymbol{n}} 
	=  \sum_{\boldsymbol{k} \in \mathbb{Z}_{\text{odd}}^{r-1}} \frac{1}{\left(\prod_{i=1}^{r-1} k_i^2\right) \left(\sum_{i=1}^{r-1} k_i\right)^2 }
	=  \sum_{\boldsymbol{k} \in \mathbb{Z}_{\text{odd}}^{r}} \frac{\delta \left(\sum_{i=1}^r k_i\right)}{\prod_{i=1}^r k_i^2}.
\end{equation}
Using the Fourier representation of the Kronecker delta function, we find
\begin{equation}
	\sum_{\boldsymbol{k} \in \mathbb{Z}_{\text{odd}}^{r}} \frac{\delta \left(\sum_{i=1}^{r} k_i\right)}{\prod_{i=1}^r k_i^2} = \int_{-\pi}^\pi \frac{dx}{2\pi} \left( \sum_{k \in \mathbb{Z}_{\text{odd}}} \frac{e^{i k x}}{k^2}\right)^r = \frac{\pi^{2r}}{4^r (r+1)} \,,
\end{equation}
where in the last step, we have used the following identity
\begin{equation}
	\sum_{k \in \mathbb{Z}_{\text{odd}}} \frac{e^{i k x}}{k^2} = \frac{\pi^2}{4} - \frac{\pi}{2} |x|  \,, \qquad x \in [-\pi, \pi]\,.
\end{equation}
We therefore get: 
\begin{equation}
	\label{eq::GE_lambda1_2}
	\lim_{\delta \to 0} G_E(q_1) = \frac{1}{m} \ln \left[ \frac{1}{2^m} + \frac{1}{2^m}   \sum_{r=1}^m \binom{m}{2r} \frac{1}{2r+1}  \right] = - \frac{1}{m} \ln (1+m).
\end{equation}
The same argument can be done for $\lambda_2$; it is easy to check that it gives the same contribution as $\lambda_1$. In Fig.~\ref{fig:Ge_Fourier} we show that the maximum value of the energetic term is always attained by the degenerate eigenvalues $\lambda_1$ and $\lambda_2$. 


\section{Generalized Reverse Wedge Perceptron}
\label{app:RWP}
\begin{figure}[htbp]
	\centering
	\begin{subfigure}[b]{0.45\textwidth}
		\includegraphics[width=\textwidth]{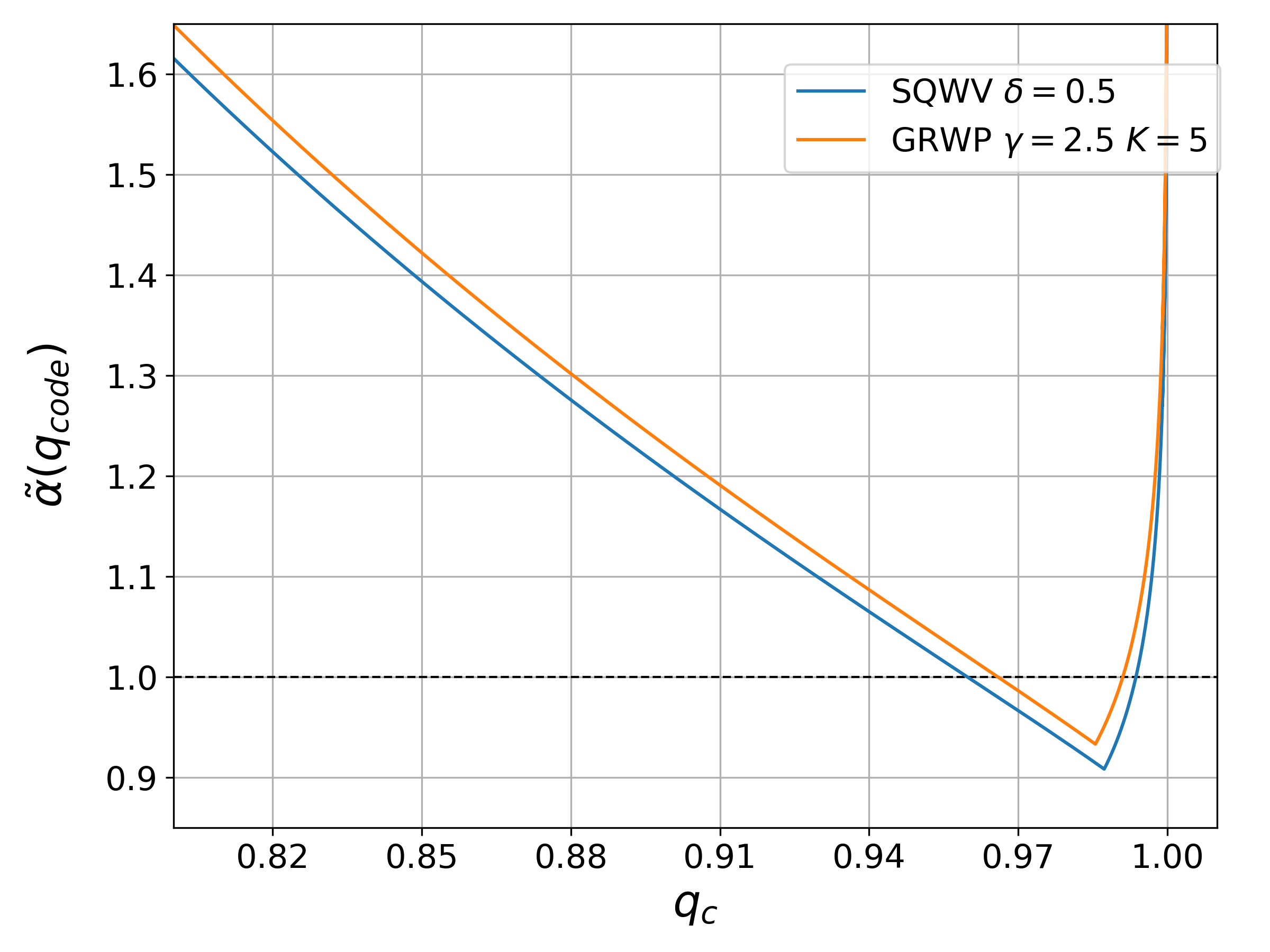}
		\caption{Comparison between the compression rate achievable using a GRWP with $K=5$ and $\gamma=2.5$ and a SWP with $\delta=0.5$. An activation with constant plateaus leads to a higher compression rate. Both activations render a CRH when composed with a code with $q_c$ in a window around $q_c=0.98$.}
		\label{fig:RWP_compression_rate}
	\end{subfigure}
	\hfill  
	\begin{subfigure}[b]{0.45\textwidth}
		\includegraphics[width=\textwidth]{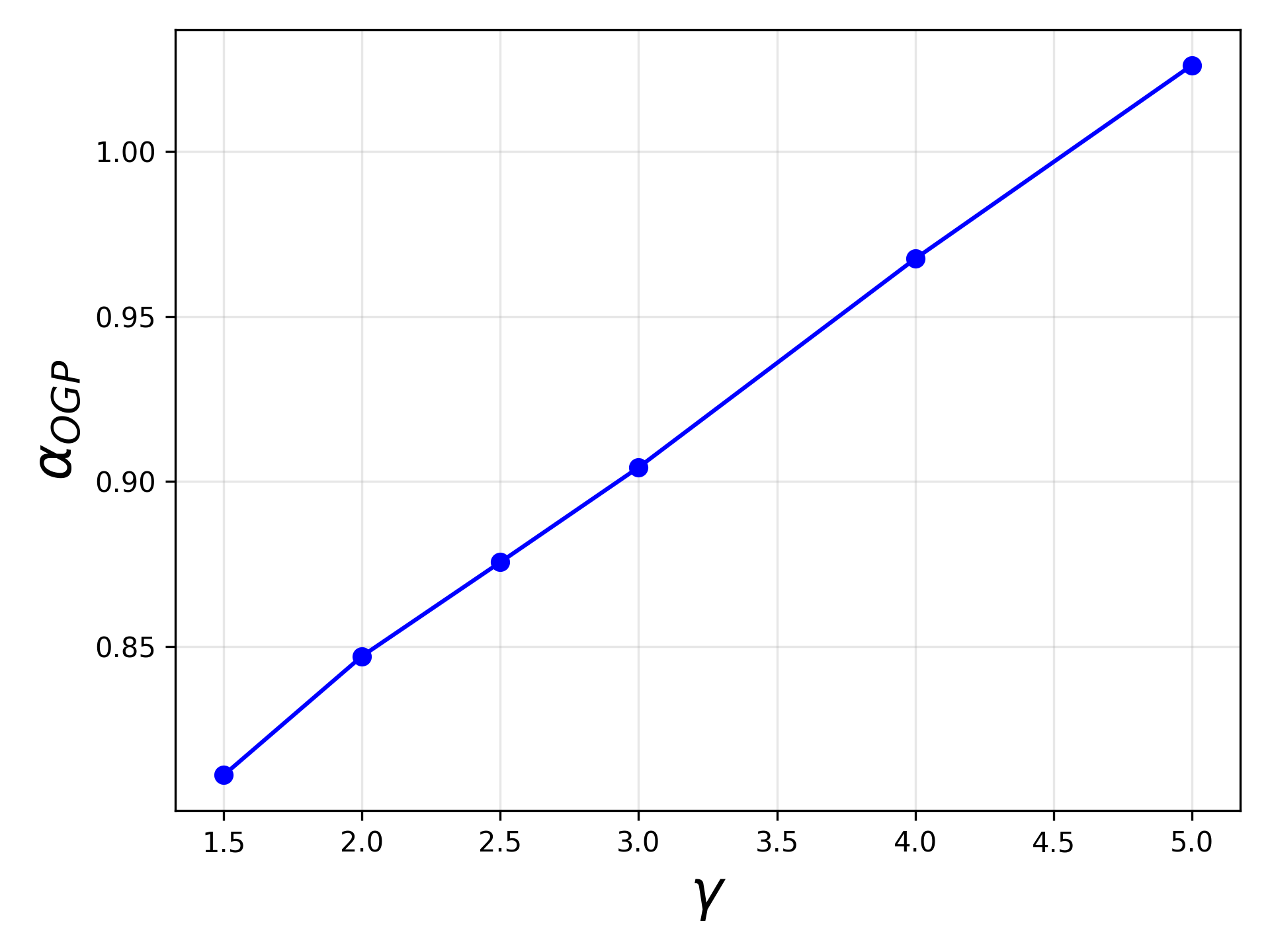}
		\caption{For the GRWP, $\alpha_{OGP}$ for the collision-finding problem for $q_1=0.05$ and $K=5$, as a function of $\gamma$. In the range of $\gamma$ we explored, decreasing $\gamma$ leads to a decrease of $\alpha_{OGP}$. Note that since $K$ fixes the number of oscillations, reducing $\gamma$ increases their frequencies.}
		\label{fig:RWP_alphaOGPvsq1}
	\end{subfigure}
\end{figure}

It is interesting to see to what extent our results rely on the activation function being periodic. We therefore applied the same formalism to the `Generalized Reverse Wedge Perceptron' (GRWP) activation function, namely a Square Wave Perceptron where oscillations are limited within a finite window. The precise shape of the activation we are considering is:
\begin{equation}
	\varphi(h) = \prod_{l=-K}^{K} \left( h + \frac{l \gamma}{K}  \right),
\end{equation}
where $2K$ is the number of oscillations within $[-\gamma, \gamma]$.  In \cref{fig:RWP_compression_rate} you can find the comparison between the compression rate achievable using a GRWP with $K=5$ and $\gamma=2.5$ and its fully periodic counterpart, namely a SWP with $\delta=0.5$. As intuitive, having an activation with constant plateaus makes finding collisions easier, leading to a higher (worse) compression rate. Still, the difference between the two curves is small, they both render a CRH when composed with a code with $q_c$ in a window around $q_c=0.98$. In \cref{fig:RWP_alphaOGPvsq1} you can find the values of $\alpha_{OGP}$ for the collision-finding problem for $q_1=0.05$ and $K=5$, as a function of $\gamma$. Notice that these low values of $q_1$ determine whether composing the network with an appropriate code will lead to a CRH or not. For the analyzed values, decreasing $\gamma$ at fixed $K$, i.e. increasing the frequency of the oscillation, leads to a decrease of $\alpha_{OGP}$. For $\gamma<2.5$ $\alpha_{OGP}$ is sufficiently low to give a CRH (compare it with the corresponding values for the SWP in Fig. 3 in the main text). \\Notice that the standard deviation of the Gaussian fields $h$ is normalized to one in our setting. Hence, $\gamma<2.5$ means that a substantial fraction of the Gaussian mass lies outside of the oscillating region of the activation function, and as a result the activation can be considered as bona fide non-periodic. This has interesting consequences in terms of comparisons with Ajtai's function, as detailed in Sec. IV.C of the main text.

\section{Algorithmic Attacks}
\subsection{Local Search From a Random Reference}
Perhaps the most straightforward strategy to look for a collision in a binary perceptron is the following. Take a random binary configuration $\boldsymbol{w}$. This trivially collides with itself. To build a non-trivial collision, the idea is to perform single bit-flips from $\boldsymbol{w}$. Here we want to count the average number $\mathcal{N}_1$ of collisions obtained by flipping one bit from the reference $\boldsymbol{w}$. This quantity is simply given by:
\begin{equation}
	\mathcal{N}_1=\sum_{i=1}^N\prod_{\mu=1}^PJ_{\varphi}^{\mu,i},
\end{equation}
where $N$ is the number of weights, $P$ is the number of examples and $p_i^{\mu}$ is the probability that if you flip bit $i$ you do not violate pattern $\mu$. One has:
\begin{equation}
	J_{\varphi}^{\mu,i}=\mathbb{E}\,\left[\Theta\left(\varphi\left(\frac{1}{\sqrt{N}}\sum_{j,j\neq i}a^{\mu}_j+\frac{1}{\sqrt{N}}a^{\mu}_i\right)\,\varphi\left(\frac{1}{\sqrt{N}}\sum_{j,j\neq i}a^{\mu}_j-\frac{1}{\sqrt{N}}a^{\mu}_i\right)\right)\right],
\end{equation}
where the $a^{\mu}_i\sim \mathcal{N}(0,1)$. Since the disorder is uncorrelated, one simply has that all the probabilities are the same, $J_{\varphi}^{\mu,i}=J_{\varphi}$, and:
\begin{equation}
	\mathcal{N}_1=N\,J_{\varphi}^P,
\end{equation}
where $J_{\varphi}$ is the same function defined in \eqref{eq:defJphi}.
Therefore, for large $N$ (see \ref{sec:AnnealedSubExt}) we find
\begin{equation}
	\mathcal{N}_1\approx N\,e^{-a_{\varphi}\,P\,N^{-1/2}},
\end{equation}
where $a_{\varphi}$ is defined in \eqref{eq:defaphi}. This means that if $P=O(N)$, in the thermodynamic limit it is not possible to move away from the initial point. 

The same computation can simply extended to the case of a fixed number $t$ of bit flips. One gets that the average number of collisions $\mathcal{N}_t$ in this case are given by: 
\begin{equation}
	\label{eq:Nt_app}
	\mathcal{N}_t\approx \binom{N}{t}\,e^{- a(\delta)\,P\,t^{1/2}\,N^{-1/2}}.
\end{equation}
Equation \eqref{eq:Nt_app} suggests that the only way to find a collision is to explore extensive distances from the reference. Indeed, if $P=O(N)$, in this regime the probability of finding a collision is exponentially small in $N$, but there is an exponential number of configurations, and the two can compensate. 

\subsection{Reinforced Approximate Message Passing}
\begin{figure}[t]
	\centering
	\begin{subfigure}{0.48\textwidth}
		\centering
		\includegraphics[width=\textwidth]{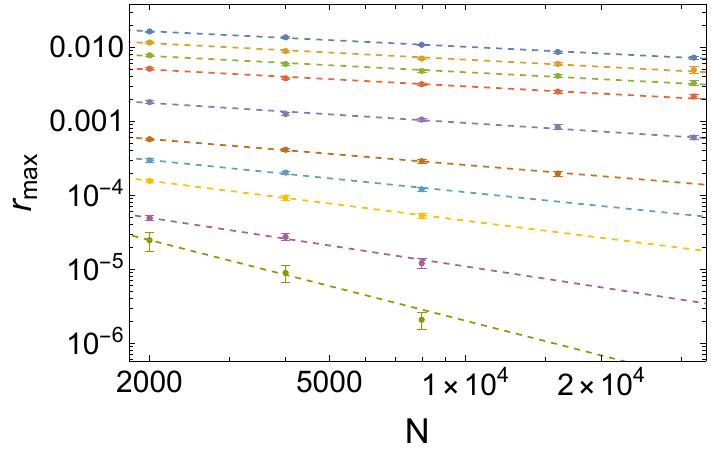}
		\caption{}
		\label{fig:rMAxvsN}
	\end{subfigure}
	\hfill
	\begin{subfigure}{0.46\textwidth}
		\centering
		\includegraphics[width=\textwidth]{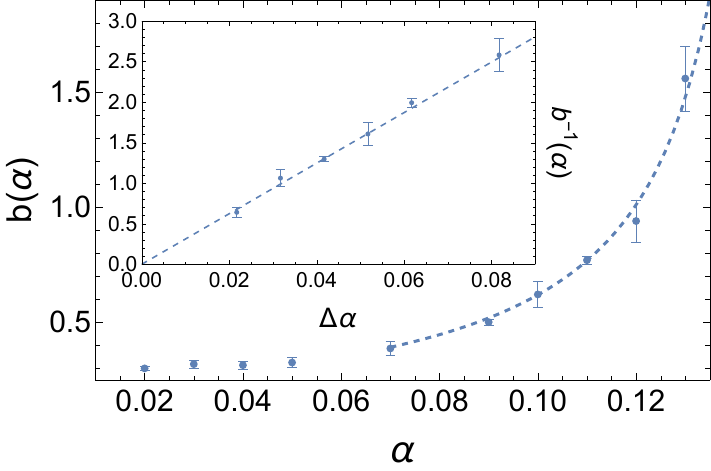}
		\caption{}
		\label{fig:balphaVSalpha}
	\end{subfigure}
	\caption{(\emph{a}): Maximum reinforcement rate $r_{\mathrm{max}}$ as a function of $N$ for different values of $\alpha$. The dashed lines are fits of the form $a(\alpha)N^{-b(\alpha)}$. Each point is obtained by averaging over $320\times 10^3/N$ independent samples. From top to bottom $\alpha$ goes from $0.02$ to $0.13$, with steps of size $0.01$. (\emph{b}): exponent $b(\alpha)$ as a function of $\alpha$. The dashed line is a fit of the form $c/(\alpha_{\mathrm{alg}}-\alpha)$, with $\alpha_{\mathrm{alg}}\approx 0.15$ and $c\approx 0.03$. \emph{Inset}: exponent $b(\alpha)$ as a function of $a-\alpha$ in log-log scale.}
	\label{fig:side_by_side1}
\end{figure}

\begin{figure}[t]
	\centering
	\begin{subfigure}{0.48\textwidth}
		\centering
		\includegraphics[width=\textwidth]{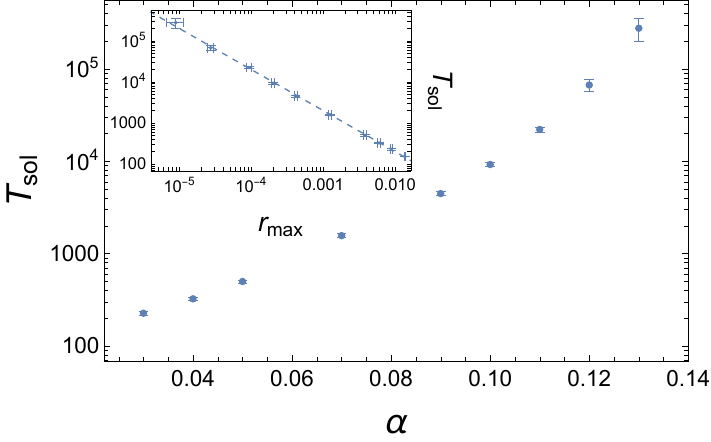}
		\caption{}
		\label{fig:Tsol}
	\end{subfigure}
	\hfill
	\begin{subfigure}{0.48\textwidth}
		\centering
		\includegraphics[width=\textwidth]{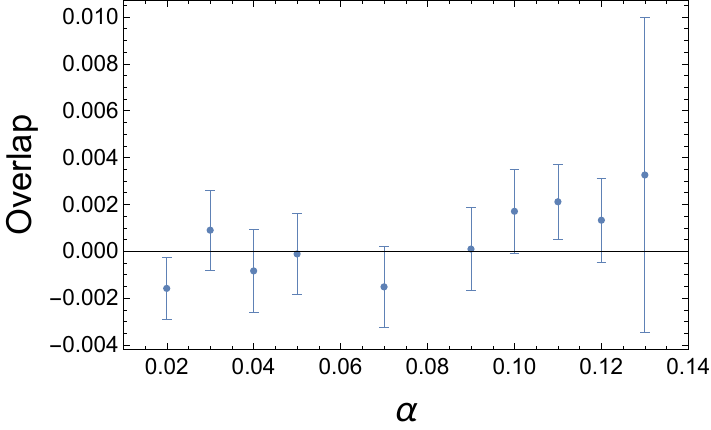}
		\caption{}
		\label{fig:overlap}
	\end{subfigure}
	\caption{(\emph{a}): number of iterations $T_{\mathrm{sol}}$ required to find a solution as a function of $\alpha$ \emph{Inset}: $T_{\mathrm{sol}}$ as a function  of $r_{\mathrm{max}}$ in log-log scale. The dashed line is proportional to $1/r_{\mathrm{max}}$. (\emph{b}): average normalized internal overlap of the collisions found with reinforcement. Data are obtained by averaging over $40$ samples of size $N=8000$.}
	\label{fig:side_by_side2}
\end{figure}

The reinforced approximate message passing algorithm (rAMP) is a message passing heuristic algorithm for finding solutions in constraint satisfaction problems. In our case, given a random vector $\boldsymbol{w}\in\{\pm 1\}^N$, we would like to find a $\boldsymbol{x}\in\{\pm 1\}^N$ such that the collision constraints
\begin{equation}
	\label{eq:CollisionConstraint}
	\varphi(\boldsymbol{A}\,\boldsymbol{w})=\varphi(\boldsymbol{A}\,\boldsymbol{x})
\end{equation}
are satisfied. We remind the reader that the disorder $\boldsymbol{A}$ is a $P\times N$ matrix, with i.i.d.\ elements distributed according to a normal with zero mean and unit variance. We call $\boldsymbol{y}$ the image of the reference $\boldsymbol{w}$, namely $\boldsymbol{y}=\varphi(\boldsymbol{A}\boldsymbol{w})$. In order to write down rAMP, let us define for each pattern $\mu=1,\dots,P$ a `latent field' $g_{\mu}$, and for each site $i=1,\dots,N$ a `magnetization' $m_i$. In the following, we use the convention that Greek indices run from $1,\dots, P$, while Latin indices form $1,\dots,N$. The rAMP equations update iteratively the magnetizations and the latent fields, starting from a random initialization $m_i^0$ and $g_{\mu}^0$. The magnetizations and latent fields at time $t$, respectively $m_i^t$ and $g_{\mu}^t$, are given by: 
\begin{subequations}
	\begin{align}
		\label{eq:AMP1}
		V_{\mu}^{t}&=\sum_i(A^{\mu}_i)^2\,\Big(1-\big(m_i^{t-1}\big)^2\Big) \\
		\label{eq:AMP2}
		\omega_{\mu}^{t}&=\sum_iA^{\mu}_i\,m_i^{t-1}-V_{\mu}^{t}\,g_{\mu}^{t-1} \\
		\label{eq:AMP3}
		g_{\mu}^{t}&=f(\omega_{\mu}^{t},V_{\mu}^{t},y_{\mu}),\quad \dot{g}_{\mu}^{t}=\partial_{\omega} f(\omega_{\mu}^{t},V_{\mu}^{t},y_{\mu}) \\
		\label{eq:AMP4}
		\Sigma_i^{t}&=\left[-\sum_{\mu}(A^{\mu}_i)^2\,\dot{g}_{\mu}^{t}\right]^{-1} \\
		\label{eq:AMP5}
		h_i^{t}&=m_i^{t-1}+\Sigma_i^{t}\,\sum_{\mu}A^{\mu}_i\,g_{\mu}^{t}+r\,t\,\tanh^{-1}(m_i^{t-1}) \\
		\label{eq:AMP6}
		m_i^{t}&=\tanh{(h_i^{t})},
	\end{align}
\end{subequations}
where the $\omega_{\mu}$, the $V_{\mu}$, the $\dot{g}_{\mu}$, $\Sigma_i$, $h_i$ are auxiliary variables defined in terms of the magnetizations and the latent fields. The function $f(\omega,V,y)$, (and its derivative $\partial_{\omega}f(\omega,V,y)$) is defined by the activation $\varphi(\bullet)$ according to the following formula:
\begin{equation}
	f(\omega,V,y)=\partial_{\omega}\log{\int dz\, e^{-\frac{(z-\omega)^2}{2V}}\Theta\big(y\,\varphi(z)\big)}.
\end{equation}
For the SWP one has:
\begin{multline}
	f(\omega,y,V)=\partial_{\omega}\log{\left(1-y\frac{4}{\pi}\sum_{n=0}^{\infty}\frac{e^{-\pi^2\,\frac{V}{\delta^2}\frac{(2n+1)^2}{2}}}{2n+1}\sin{\left(\frac{\pi}{\delta}(2n+1)\,\omega\right)}\right)}=\\= -y\frac{4}{\delta}\frac{\sum_{n=0}^{\infty}e^{-\pi^2\,\frac{V}{\delta^2}\frac{(2n+1)^2}{2}}\cos{\left(\frac{\pi}{\delta}(2n+1)\,\omega\right)}}{1-y\frac{4}{\pi}\sum_{n=0}^{\infty}\frac{e^{-\pi^2\,\frac{V}{\delta^2}\frac{(2n+1)^2}{2}}}{2n+1}\sin{\left(\frac{\pi}{\delta}(2n+1)\,\omega\right)}}.
\end{multline}
\begin{equation}
	\partial_{\omega}f(\omega,y,V)=-f^2(\omega,y,V)+y\frac{4\pi}{\delta^2}\frac{\sum_{n=0}^{\infty}e^{-\pi^2\,\frac{V}{\delta^2}\frac{(2n+1)^2}{2}}(2n+1)\sin{\left(\frac{\pi}{\delta}(2n+1)\,\omega\right)}}{1-y\frac{4}{\pi}\sum_{n=0}^{\infty}\frac{e^{-\pi^2\,\frac{V}{\delta^2}\frac{(2n+1)^2}{2}}}{2n+1}\sin{\left(\frac{\pi}{\delta}(2n+1)\,\omega\right)}}.
\end{equation}
The parameter $r$ in \eqref{eq:AMP5} is called reinforcement rate. Setting $r=0$, the rAMP becomes equivalent to the approximate message passing (AMP) algorithm, which is used to estimate the local marginals of variables in constraint-satisfaction problems. Setting $r>0$ one obtains a heuristic solver \cite{chavas2005survey,braunstein2006learning}. Indeed, the reinforcement term can be interpreted as a soft decimation that at long times forces the magnetizations to polarize. For $r>0$, equations (\ref{eq:AMP1})-(\ref{eq:AMP6}) are iterated until the binary vector $\tilde{\boldsymbol{x}}^t$, with elements $\tilde{x}^t_i=\text{sign}(m_i^t)$, satisfies constraint \ref{eq:CollisionConstraint}, or $t$ becomes bigger that a maximum time $T_{\mathrm{max}}$, that is a parameter of the algorithm. There are different choices of the reinforcement term suggested in the literature, which all lead to qualitatively similar results. The behavior of the solver depends on the choice of $r$.  One finds that the performance of the solver, which is determined by the largest value of $\alpha$ at which the solver is able to find solutions in polynomial time, improves the lower the value of $r$. However, it is also found that the number of iterations required to find a solution scales like $1/r$ \cite{braunstein2006learning,baldassi2015max}. Therefore, it is convenient to define the maximum reinforcement rate $r_{\mathrm{max}}$, which is the largest reinforcement rate allowing to find a solution. The quantity $r_{\mathrm{max}}$ depends on $\alpha$, on the size $N$, and it fluctuates from sample to sample. From numerical simulations it is found that $r_{\mathrm{max}}\approx a(\alpha)N^{-b(\alpha)}$ (see Fig.~\ref{fig:rMAxvsN}). The exponent $b(\alpha)$ grows approaching $\alpha_{\mathrm{alg}}\approx 0.15$, compatibly with $(\alpha_{\mathrm{alg}}-\alpha)^{-1}$ (see Fig.~\ref{fig:balphaVSalpha}). The exponent $b(\alpha)$ is a crucial quantity for the efficiency of the solver. Indeed, note that the rAMP equations (\ref{eq:AMP1})-(\ref{eq:AMP6}) involve matrix vector multiplications, that require $O(NP)$ operations. Since $P=O(N)$, each iteration of rAMP takes $O(N^2)$ operations. The total complexity of the algorithm also depends on the number of iterations $T_{\mathrm{sol}}$ required to find a solution. From numerical simulations, $T_{\mathrm{sol}}\approx r_{\mathrm{max}}^{-1}$ (see Fig.~\ref{fig:Tsol}). Therefore, the complexity at a given $\alpha$ is proportional to $N^{2+b(\alpha)}$. In Fig.~\ref{fig:overlap} we plot the average internal overlap $\boldsymbol{w}\cdot\boldsymbol{x}/N$, between $\boldsymbol{w}$ and the $\boldsymbol{x}$ returned by reinforcement. Interestingly, the solutions $\boldsymbol{x}$ found by rAMP are almost orthogonal to the reference $\boldsymbol{w}$. 

We conclude noting that the gap between the OGP estimate for $q_1=0$, that for $m=5$ and $\delta=0.6$ is $\alpha_{OGP}^{5}\approx 0.9$, is definitely larger than the estimated algorithmic threshold for rAMP, $\alpha_{\mathrm{alg}}=0.15$. Still, it should be stressed that: i) the OGP threshold we compute is an upper bound on the true one; ii) we carried out the computation up to the $5$-OGP at finite $\delta$, but higher values of $m$ should yield lower thresholds; iii) the algorithmic attack we consider is of the second-preimage type: one weight vector is chosen at random, and the other is found via rAMP. This setup is equivalent to the teacher–student problem (TSP). Note that the TSP is harder than collisions (a solver for TSP automatically finds collisions, but not vice versa). In \cite{benedetti2025overlap}, OGP thresholds are computed on the TSP for various values of $\delta$, also at the RS level. For $\delta = 0.6$, the best OGP estimate is $\approx 0.25$—much closer to $\alpha_{\mathrm{alg}}=0.15$, as one would expect.


\section{Exhaustive search measure of 2-OGP}
\label{app:Exhaustive_search}
\begin{figure}[h]
	\centering
	\begin{subfigure}{0.45\textwidth}
		\centering
		\includegraphics[trim={4cm 0cm 30.7cm 3cm}, clip, width=\textwidth]{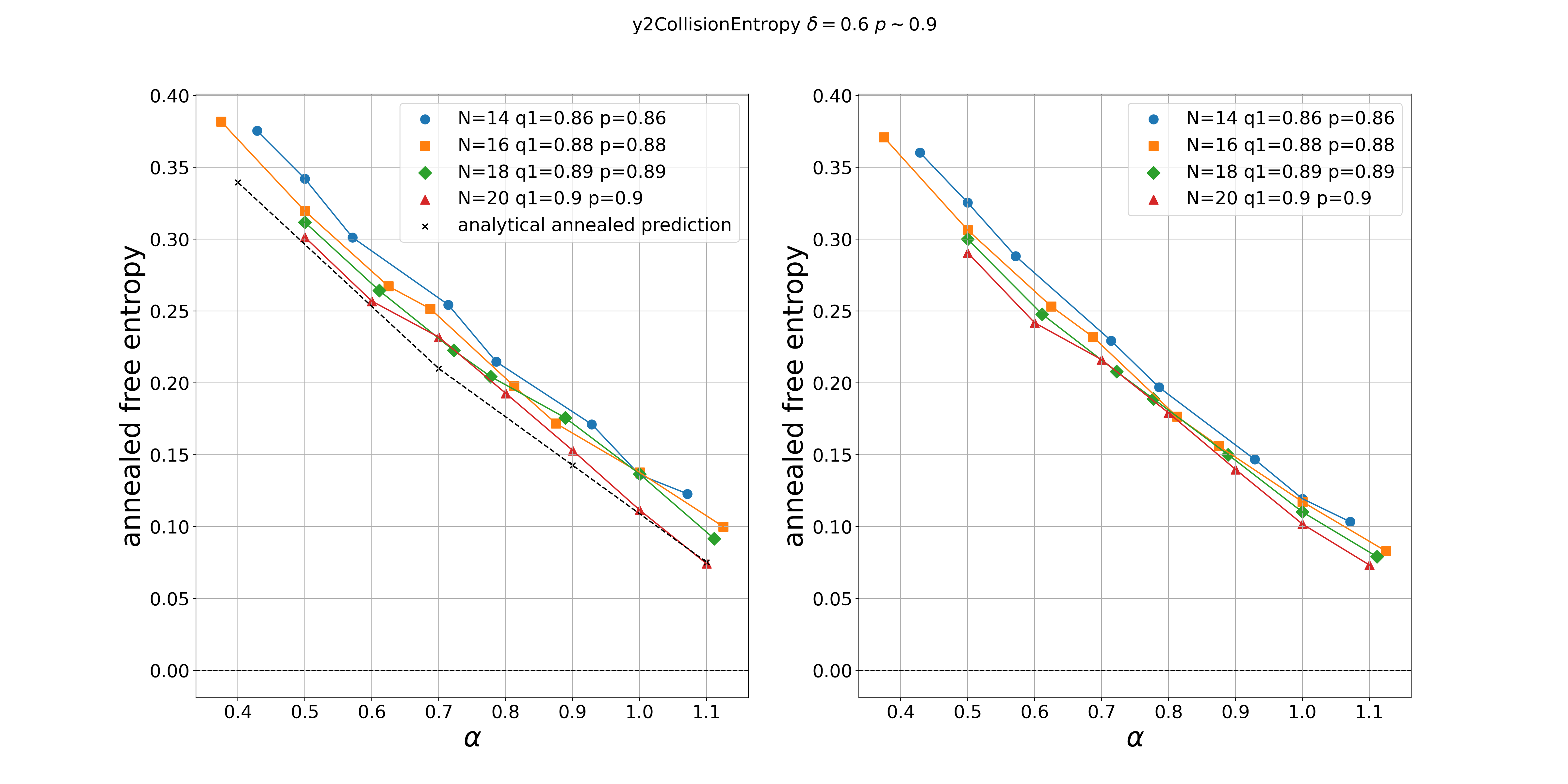}
	\end{subfigure}
	\begin{subfigure}{0.45\textwidth}
		\centering
		\includegraphics[trim={4cm 0 30.7cm 3cm}, clip, width=\textwidth]{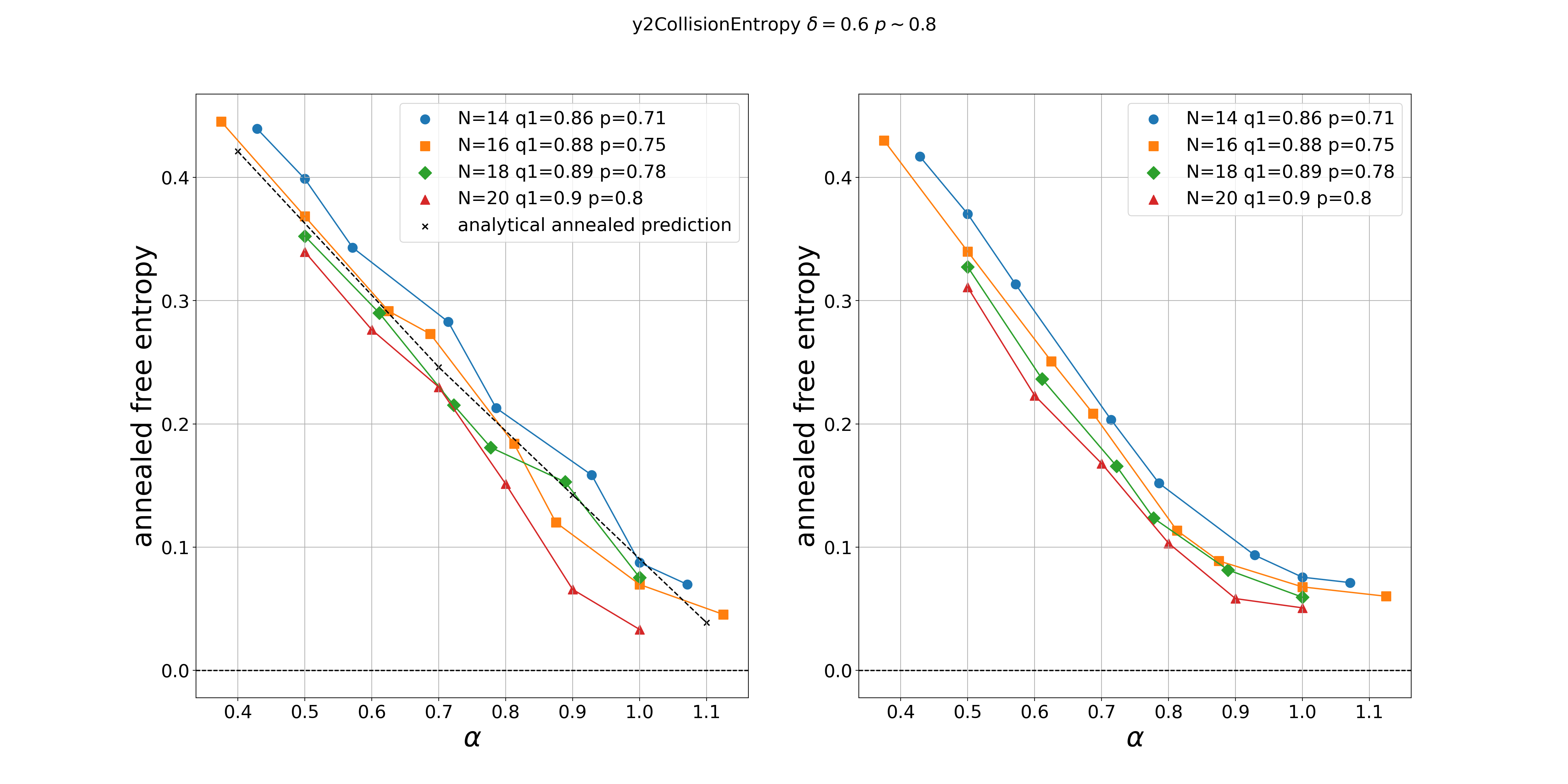}
	\end{subfigure}
	\caption{Exhaustive search estimates and analytical predictions of the annealed free entropy for $q_1=0.9$, and various values of $N$. The dashed line represents the annealed analytical prediction, for $p=0.9$ (left) and $p=0.75$ (right).  Full lines represent the numerical exhaustive search estimate, for the closest measurable values of $p$.}
	\label{fig:2OGP_exaustive}
\end{figure}
The analytical prediction for the annealed free entropy $\phi_m^{\mathrm{a}}(d) = \lim_{N \to \infty} \frac{1}{mN} \log \mathbb{E}_{\boldsymbol{A}} \mathcal{N}_m(p, q_1; \boldsymbol{A})$ can be tested numerically. Recall that $d=1-p = 1-\frac{1}{2N}\big(\langle \vecx_i, \vecx_j\rangle +\langle \vecy_i, \vecy_j \rangle \big)$ is the distance between collision $i$ and $j$, and we are considering only collisions with a given internal overlap $q_1=\langle \vecx_1, \vecy_1\rangle=...= \langle \vecx_m; \vecy_m\rangle$.\\

To do so, one generates a realization of disorder $\matA$, and iterates through all $2^N$ weights, grouping all those generating the same output $y$. Within each group of weights, one iterates through all 2-uples, selecting only those such that $\langle \vecx, \vecy\rangle=q_1$. Finally, one must iterate through all m-uples of such collisions, and select only those such that $d(c^a;c^b)=d$. The procedure is extremely expensive from the computational point of view, thus the analysis will be limited to:
\begin{itemize}
	\item the case $m=2$;
	\item small values of $N=14, 16, 18, 20$;
	\item the highest value of $q_1$ measurable for each $N$, namely $(N-2)/N$;
	\item high values of $p$, namely $(2N-4)/N,\,(2N-8)/N$, for each $N$.
\end{itemize}
\Cref{fig:2OGP_exaustive} present the numerical estimates of $\phi_m^{\mathrm{a}}(d)$ for different values of $N$, as a function of $\alpha$, for the two measured values of $p$. The dashed line represents the prediction from the analytical computation, where the values of the overlaps are fixed to $q_1=0.9,\,p=0.9$ and $q_1=0.9,\,p=0.75$ respectively. In comparison with the numerical results, we can see that the matching with theory is very good. One might be worried by the fact that increasing $N$ seems to systematically lower the numerical curves. In fact, this cannot be directly attributed to finite-size effects, but rather to the fact that the values of $q_1$ and $p$ are slightly different among the curves, as reported in the key.\\

\end{document}